\documentclass[journal,twoside,twocolumn,a4paper]{IEEEtran}
\usepackage{amsmath,amssymb,amscd,latexsym,dsfont}
\usepackage{multirow}
\usepackage{float}
\usepackage{comment}
\usepackage{color}
\usepackage{graphicx}
\usepackage{comment}
\usepackage{subfigure}
\usepackage{enumerate}
\usepackage{stfloats}
\usepackage{psfrag}
\usepackage{setspace}
\usepackage{cite}
\usepackage{balance}
\usepackage{algorithmic}
\usepackage[]{algorithm}
\usepackage{xcolor}

\newcommand{\lefto}{\mathopen{}\left} %
\newcommand{\abs}[1]{\left\lvert#1\right\rvert}		%

\graphicspath{{figures/}{figures/contours/}}
\IEEEoverridecommandlockouts
\newcommand{\eqlab}[2]{\begin{align} \label{#1} #2 \end{align}}
\newcommand{\eq}[1]{\begin{align*} #1 \end{align*}}
\newcommand{\E}{\mathbb{E}}
\newcommand{\pdf}{f}

\newcommand{\sumi}{\sum_{i=1}^{16}}

\newcommand{\suml}{\sum_{l=0}^{4N}}

\newcommand{\tr}[1]{\mathrm{#1}}
\newcommand{\tnr}[1]{{\textnormal{#1}}}

\newcommand{\mc}[1]{\mathcal{#1}}

\newcommand{\set}[1]{\{#1\}}
\newcommand{\cd}{\cdot}

\newcommand{\ld}{\ldots}

\newcommand{\ms}[1]{\mathds{#1}}

\newcommand{\ov}[1]{\overline{#1}}

\newcommand{\nchoosek}[2]{\binom{#1}{#2}}

\newcommand{\ie}{i.e.,~}

\newtheorem{theorem}{Theorem}
\newcounter{lemma}
\newtheorem{corollary}[lemma]{Corollary}
\newtheorem{lemma}{Lemma}

\newtheorem{remark}{Remark}

\newcommand{\QF}[0]{Q}

\newcommand{\bX}{\boldsymbol{X}}

\newcommand{\bZ}{\boldsymbol{Z}}

\newcommand{\bY}{\boldsymbol{Y}}

\newcommand{\mcI}{\mc{I}}

\newcommand{\mcS}{\mc{S}}

\newcommand{\mcV}{\mc{V}}

\newcommand{\bU}{\boldsymbol{U}}

\newcommand{\bx}{\boldsymbol{x}}

\newcommand{\bu}{\boldsymbol{u}}

\newcommand{\by}{\boldsymbol{y}}

\newcommand{\argmin}{\mathop{\mathrm{argmin}}}

\newcommand{\Px}{P}

\newcommand{\Ex}{\ms{E}}
\renewcommand{\exp}{\mathrm{exp}}

\newcommand{\figref}[1]{Fig.~\ref{#1}}
\newcommand{\tabref}[1]{Table~\ref{#1}}
\newcommand{\secref}[1]{Sec.~\ref{#1}}
\newcommand{\theoref}[1]{Theorem~\ref{#1}}
\newcommand{\theosref}[2]{Theorems~\ref{#1} and \ref{#2}}

\newcommand{\appref}[1]{Appendix~\ref{#1}}

\newcommand{\Yn}{Y_{k}}

\newcommand{\Xn}{X_{k}}\newcommand{\xn}{x_{k}}
\newcommand{\Zn}{Z_{k}}

\newcommand{\X}{X}
\newcommand{\hX}{\hat{X}}

\newcommand{\XkmN}{X_{k-N}}

\newcommand{\Xkmo}{X_{k-1}}

\newcommand{\Xkpo}{X_{k+1}}

\newcommand{\XkpN}{X_{k+N}}

\newcommand{\Y}{Y}
\newcommand{\y}{y}

\newcommand{\Szsilower}{\rho(\abs{s_i}^2\!+\|\xmem\|^2)}%
\newcommand{\Szsilowerl}{\rho(|s_{i}|^2+\delta_{l})}

\newcommand{\Znt}{\tilde{Z}_{k}}

\newcommand{\gammaXNl}{\gamma_{l,t,N}}
\newcommand{\gammaltN}[3]{\gamma_{#1,#2,#3}}

\newcommand{\Xmem}{\bX^{\text{mem}}_k}%
\newcommand{\xmem}{\bx^{\text{mem}}_k}%

\newcommand{\PASE}{P_\text{ASE}}

\newcommand{\SEP}{\tr{SER}}
\newcommand{\BEP}{\tr{BER}}

\newcommand{\reals}{\mathbb{R}}

\newcommand{\header}{}
\title{Capacity of a Nonlinear Optical Channel\\with Finite Memory}
\author{Erik Agrell, Alex Alvarado, Giuseppe Durisi, and Magnus Karlsson
\thanks{Research supported by the Swedish Research Council (VR) under grant no.~2012-5280, the Swedish Foundation for Strategic Research (SSF) under grant no.~RE07-0026, and the European Community's Seventh's Framework Programme (FP7/2007-2013) under grant agreement no.~271986. This paper was presented in part at the 2013 European Conference on Optical Communication.}
\thanks{E.~Agrell and G.~Durisi are with the Department of Signals and Systems, Chalmers University of Technology, SE-41296 Gothenburg, Sweden (email: 
\{agrell,durisi\}@chalmers.se).}
\thanks{M.~Karlsson is with the Department of Microtechnology and Nanoscience, Chalmers University of Technology, SE-41296 Gothenburg, Sweden (email: 
magnus.karlsson@chalmers.se).}
\thanks{A.~Alvarado is with the Department of Engineering, University of Cambridge, Cambridge CB2~1PZ, United Kingdom (email: alex.alvarado@ieee.org).}
}
\begin{document}
\maketitle

\begin{abstract}
The channel capacity of a nonlinear, dispersive fiber-optic link is revisited. To this end, the popular Gaussian noise (GN) model is extended with a parameter to account for the finite memory of realistic fiber channels. This finite-memory model is harder to analyze mathematically but, in contrast to previous models, it is valid also for nonstationary or heavy-tailed input signals. For uncoded transmission and standard modulation formats, the new model gives the same results as the regular GN model when the memory of the channel is about 10 symbols or more. These results confirm previous results that the GN model is accurate for uncoded transmission. However, when coding is considered, the results obtained using the finite-memory model are very different from those obtained by previous models, even when the channel memory is large. In particular, the peaky behavior of the channel capacity, which has been reported for numerous nonlinear channel models, appears to be an artifact of applying models derived for independent input in a coded (i.e., dependent) scenario.
\end{abstract}


\section{Introduction}\label{Sec:Introduction}

The introduction of coherent optical receivers has brought significant advantages in fiber optical communications, e.g., enabling efficient polarization demultiplexing, higher-order modulation formats, increased sensitivity, and electrical mitigation of transmission impairments\cite{sun08,roberts09}. Even if the linear transmission impairments (such as chromatic and polarization-mode dispersion) can be dealt with electronically, the Kerr nonlinearity in the fiber remains a significant obstacle. Since the nonlinearity causes signal distortions at high signaling powers, arbitrarily high signal-to-noise ratios are inaccessible, which limits transmission over long distances and high spectral efficiencies. This is sometimes referred to as the ``nonlinear Shannon limit'' \cite{Ellis10,Mecozzi12}.

For systems with large accumulated dispersion and weak nonlinearity, the joint effect of chromatic dispersion and the Kerr effect is similar to that of additive Gaussian noise. This was pointed out already by Splett \cite{splett93} and Tang \cite{tang02}. The emergence of this Gaussian noise is prevalent in links that have no inline dispersion compensation, such as today's  coherent links, where the dispersion compensation takes place electronically in the receiver signal processing. 
This Gaussian noise approximation has been recently rediscovered and applied to today's coherent links in a series of papers by Poggiolini \emph{et al.} \cite{poggiolini11,bosco11,carena12,poggiolini12} and other groups \cite{beygi12,Johannisson13,grellier11}. 
The resulting so-called Gaussian noise model, or \emph{GN model} for short, is  valid for multi-channel (wavelength- and polarization-division multiplexed) signals. 
It has also been shown to work for single-channel and single-polarization transmission if the dispersive decorrelation is large enough \cite{beygi12,beygi13}. %

A crucial assumption in the derivation of the GN model is that of independent, identically distributed (i.i.d.) inputs: the transmitted symbols are independent of each other, are drawn from the same constellation, and have the same constellation scaling (the same average transmit power).
Under these assumptions, the model has been experimentally verified to be very accurate \cite{vacondio12,carena12} for the most common modulation formats, such as quadrature amplitude modulation (QAM) or phase-shift keying.

In this paper, the assumption of i.i.d.~inputs is, perhaps for the first time in optical channel modeling, relaxed. This is done by introducing a modified GN model, which we call the \emph{finite-memory GN model}. This new model includes the memory of the channel as a parameter and differs from previous channel models in that it is valid also when the channel input statistics are time-varying, or when ``heavy-tailed'' constellations are used.

The performance predicted by the regular GN model (both in terms of uncoded error probability and channel capacity) is compared with the ones predicted by the finite-memory GN model.
The \emph{uncoded} performance is characterized in terms of symbol error rate (SER) and bit error rate (BER), assuming i.i.d.~data. Exact analytical expressions are obtained for 16-ary QAM (16-QAM), which show that the GN model is accurate for uncoded transmission and standard modulation formats, confirming previous results.

The main contributions of the paper are in terms of \emph{coded} performance. Shannon, the father of information theory, proved that for a given channel, it is possible to achieve an arbitrarily small error probability, if the transmission rate in bits per symbol is small enough. A rate for which virtually error-free transmission is possible is called an \emph{achievable rate} and the supremum over all achievable rates for a given channel, represented as a statistical relation between its input $X$ and output $Y$, is defined as the \emph{channel capacity} \cite{shannon48}, \cite[p.~195]{cover06}.
A capacity-approaching transmission scheme operates in general by grouping the data to be transmitted into blocks, encoding each block into a sequence of coded symbols, modulating and transmitting this sequence over the channel, and decoding the block in the receiver. This coding process introduces, by definition, dependencies among the transmitted symbols, which is the reason why channel models derived for i.i.d.~inputs may be questionable for the purpose of capacity analysis. %

More fundamentally, the regular GN model is not well-suited to capacity analysis, because in this model each output sample depends on the statistics of the previously transmitted input symbols (through their average power) rather than on their actual value.
This yields artifacts in capacity analysis. 
One such artifact is the peaky behavior of the capacity of the GN model as a function of the transmit power.
Indeed, through a capacity lower bound it is shown in this paper that this peaky behavior does not occur for the finite-memory GN model, even when the memory is taken to be arbitrary large.

The analysis of channel capacity for fiber-optical transmission dates back to 1993 \cite{splett93}, when Splett \emph{et al.} quantified the impact of nonlinear four-wave mixing on the channel capacity. 
By applying Shannon's formula for the additive white Gaussian noise (AWGN) channel capacity to a channel with power-dependent noise, Splett \emph{et al.} found that there exists an ``optimal'' \emph{finite} signal-to-noise ratio that maximizes capacity. Beyond this value, capacity starts decreasing. 
It was however not motivated in \cite{splett93} why the noise was assumed Gaussian. 
Using a different model for four-wave mixing, Stark \cite{Stark99}  showed that capacity saturates, but does not decrease, at high power. 
In the same paper, the capacity loss due to the quantum nature of light was quantified. In 2001, Mitra and Stark \cite{mitra01} considered the capacity in links where cross-phase modulation dominates, proved that the capacity is lower-bounded by the capacity of a linear, Gaussian channel with the same input--output covariance matrix, and evaluated this bound via Shannon's AWGN formula. The obtained bound vanishes at high input power. It was claimed, without motivation, that the true capacity would have the same qualitative nonmonotonic behavior.

Since 2001, the interest in optical channel capacity has virtually exploded. The zero-dispersion channel was considered by Turitsyn \emph{et al}. \cite{turitsyn03}. The joint effect of nonlinearity and dispersion was modeled by Djordjevic \emph{et al.} \cite{djordjevic05} as a finite-state machine, which allowed the capacity to be estimated using the Bahl--Cocke--Jelinek--Raviv (BCJR) algorithm. Taghavi \emph{et al.} \cite{taghavi06} considered a fiber-optical multiuser system as a multiple-access channel and characterized its capacity region. In a very detailed tutorial paper, Essiambre \emph{et al.} \cite{Essiambre10} applied a channel model based on extensive lookup tables and obtained capacity lower bounds for a variety of scenarios. Secondini \emph{et al.} \cite{secondini13} obtained lower bounds using the theory of mismatched decoding. Recently, Dar \emph{et al.} \cite{Dar14} modeled the nonlinear phase noise as being blockwise constant for a certain number of symbols, which is a channel with finite memory, obtaining improved capacity bounds.

Detailed literature reviews are provided in \cite{Kahn04} for the early results, and in \cite{Essiambre10} for more recent results. Other capacity estimates, or lower bounds thereon, were reported for various nonlinear transmission scenarios in, e.g., \cite{Narimanov02, wegener04, essiambre08, freckmann09, djordjevic10, killey11, bosco11, Mecozzi12}. Most of these estimates or bounds decrease to zero as the power increases.

This paper is organized as follows. In \secref{Sec:Model}, the GN model is reviewed and the finite-memory GN model is introduced. In \secref{Sec:Error}, the uncoded error performance of the new finite-memory model is studied. The channel capacity is studied in \secref{Sec:Capacity} and  conclusions are drawn in \secref{Sec:Conclusions}. The mathematical proofs are relegated to appendices.

\emph{Notation:}
Throughout this paper, vectors are denoted by boldface letters $\bx$ and sets are denoted by calligraphic letters $\mc{X}$. Random variables are denoted by uppercase letters $X$ and their (deterministic) outcomes by the same letter in lowercase $x$. Probability density functions (PDFs) and conditional PDFs are denoted by $f_{Y}(y)$ and $f_{Y|X}(y|x)$, respectively. Analogously, probability mass functions (PMF) are denoted by $P_{X}(x)$ and $P_{X|Y}(x|y)$. Expectations are denoted by $\E[\cd]$ and random sequences by $\{{Z}_k\}$.

\section{Channel Modeling: Finite and Infinite Memory}\label{Sec:Model}

In this section, we will begin with a high-level description of the nonlinear interference in optical dual-polarization wavelength-division multiplexing (WDM) systems, highlighting the role of the channel memory, and thereafter in \secref{sec:finite_memory}--\ref{Sec:Model.Finite} describe in detail the channel models considered in this paper.

\subsection{Nonlinear Interference in  Optical Channels} %
\label{sec:nonlinear_interference_in_optical_channels}

A coherent optical communication link converts a discrete, complex-valued electric data signal $x_k$ to a modulated, continuous optical signal, which is transmitted through an optical fiber, received coherently, and then converted back to a discrete output sequence $Y_k$. 
The coherent link is particularly simple theoretically, in that the transmitter and receiver directly map the electric data to the optical \emph{field}, which is a linear operation (in contrast with, e.g., direct-detection receivers), and can ideally be performed without distortions. The channel is then well described by the propagation of the (continuous) optical field in the fiber link. 
It should be emphasized that this assumes the coherent receiver to be ideal, with perfect synchronization and negligible phase noise. 
Experiments have shown \cite{roberts09} that commercial coherent receivers can indeed perform well enough for the fiber propagation effects to be the main limitations. 
Two main \emph{linear} propagation effects in the fiber need to be addressed: \emph{dispersion} and \emph{attenuation}. 
The attenuation effects can be overcome by periodic optical amplification, at the expense of additive Gaussian noise from the inline amplifiers. 
The dispersion effects are usually equalized electronically by a filter in the coherent receiver.  Such a linear optical link can be well-described by an AWGN channel, the capacity of which is unbounded with the signal power. 

However, the fiber Kerr-nonlinearity introduces signal distortions, and greatly complicates the transmission modeling. The nonlinear signal propagation in the fiber is described by a nonlinear partial differential equation, the \emph{nonlinear Schr\"odinger equation} (NLSE), which includes dispersion, attenuation, and nonlinearity. At high power levels, the three effects can no longer be conveniently separated. 
However, in contemporary coherent links (distance at least $500~\tnr{km}$ and symbol rate at least $28~\tnr{Gbaud}$), the nonlinearity is significantly weaker than the other two effects, and a  perturbation approach can be successfully applied to the NLSE \cite{splett93,poggiolini12, beygi12,Johannisson13}. This leads to the GN model, which will be described in \secref{Sec:Model.Infinite}.

\subsection{Finite Memory} %
\label{sec:finite_memory}

Even today's highly dispersive optical links have a finite memory. For example, a signal with dispersive length $L_{\text{D}}=1/(\Delta \omega^2|\beta_2|)$, where $\beta_2$ is the group velocity dispersion and $\Delta \omega$ the optical bandwidth, broadens (temporally) a factor $L/L_{\text{D}}$ over a fiber of length $L$. With typical dispersion lengths of $5$--$50~\tnr{km}$, this broadening factor can correspond to hundreds to thousands of adjacent symbols,  a large but \emph{finite} number. 
The same will hold for interaction among WDM channels; if one interprets $\Delta \omega$ as the channel separation, $L/L_{\text{D}}$ will give an approximation of the number of symbols that two WDM channels separate due to walk-off (and hence interact  with nonlinearly during transmission). The channel memory will thus be even larger in the WDM case, and increase with channel separation, but the nonlinear interaction will decrease due to the shorter $L_{\text{D}}$. 
Thus, the principle of a finite channel memory holds also for WDM signals. To keep notation as simple as possible, we will consider a single, scalar, wavelength channel in this paper. Extensions to dual polarizations and WDM are possible, but will involve obscuring complications such as four-dimensional constellation space \cite{agrell09} in the former case and behavioral models \cite{agrellofc13} in the latter.
We can thus say that in an optical link a certain signal may sense the interference from $N\approx L/L_{\text{D}}$ neighboring symbols, which is the physical reason for introducing a finite-memory model.

If we let the range $N$ of the interfering symbols go to infinity, an even simpler type of model is obtained. The interference is now averaged over infinitely many transmitted symbols. Assuming that an i.i.d. sequence is transmitted, this time average converges to a statistical average, which greatly simplifies the analysis. All models suggested for dispersive optical channels so far belong to this category \cite{splett93, Essiambre10, Goebel11, poggiolini12, Mecozzi12, beygi12,Johannisson13,beygi13}, of which the GN model described in \secref{Sec:Model.Infinite} is the most common.

For a given transmitted complex symbol $\xn$, the (complex) single-channel output at each discrete-time $k\in\mathbb{Z}$ is modeled as
\eqlab{eq:awgn}{
\Yn = \xn + \Zn
,}
where $\{\Zn\}$ is a circularly symmetric, complex, white, Gaussian random sequence, independent of $\xn$. In \eqref{eq:awgn}, $\Zn$ is assumed to be independent of the actual transmitted sequence $\xn$. However, the \emph{variance} of $\Zn$ depends on the transmit power, as detailed in \secref{Sec:Model.Infinite} and \ref{Sec:Model.Finite}.

\subsection{The Regular GN Model}\label{Sec:Model.Infinite}

For coherent long-haul fiber-optical links without dispersion compensation, Splett \emph{et al.} \cite{splett93}, Poggiolini \emph{et al.} \cite{poggiolini11}, and Beygi \emph{et al.} \cite{beygi12} have all derived models where the nonlinear interference (NLI) appears as Gaussian noise, whose statistics depend on the transmitted signal power via a cubic relationship. The models assume that the transmitted symbols $x_k$ in time slot $k \in \mathbb{Z}$ are i.i.d.\ realizations of the same complex random variable $X$. In this model, the additive noise in \eqref{eq:awgn} is given by
\begin{align}
\label{infinite.memory.IO}
Z_k = \tilde{Z}_k \sqrt{\PASE + \eta \Px^3}
\end{align}
where $\{\tilde{Z}_k\}$ are i.i.d.\ zero-mean unit-variance circularly symmetric complex Gaussian random variables, $\PASE$ and $\eta$ are real, nonnegative constants, and $\Px = \E[|X|^2]$ is the \emph{average transmit power.} Therefore, the noise $Z_k$ is distributed as $Z_{k}\sim\mc{CN}(0,\PASE + \eta \Px^3)$, where $\mc{CN}(0,\sigma^2)$ denotes a circularly symmetric complex Gaussian random variable with mean $0$ and variance $\sigma^2$. The parameter $P$, which is a property of the transmitter, governs the behavior of the channel model. This can be intuitively understood as a long-term average of the input power. Mathematically, 
\begin{align}\label{P.statistical}
P = \lim_{N\rightarrow\infty} \frac{1}{2N+1} \sum_{i=k-N}^{k+N} |x_i|^2
\end{align}
for any given $k$, still assuming i.i.d.~symbols $x_k$.
For this reason, we will refer to models that depend on infinitely many past and/or future symbols, via $P$ in \eqref{P.statistical} or in some other way, as \emph{infinite-memory} models.

The cubic relation in~\eqref{infinite.memory.IO} between the transmit power and the additive noise variance $\PASE + \eta \Px^3$ is a consequence of the Kerr nonlinearity, and holds for both lumped and distributed amplification schemes. The constant $\PASE$ represents the total amplified spontaneous emission (ASE) noise of the optical amplifiers for the channel under study, while $\eta$ quantifies the NLI. Several related expressions for this coefficient have been proposed. For example, for distributed amplification and WDM signaling over the length $L$,
\eqlab{nli:splett}{
\eta &= \frac{4\gamma^2L}{\pi |\beta_2| B^2}\log_e\left(2\pi e |\beta_2| L B^2\right), \\
\eta &= \frac{16\gamma^2L}{27\pi |\beta_2| R_\text{s}^2}\log_e\left(\frac{2}{3}\pi^2 |\beta_2| L B^2\right), \label{nli:bosco}
}
were proposed in \cite{splett93} and \cite{bosco12}, resp., where $\gamma$ is the fiber nonlinear coefficient, $B$ is the total WDM bandwidth, and $R_\text{s}$ is the symbol rate. Obviously, the expressions in \eqref{nli:splett} and \eqref{nli:bosco} are qualitatively similar. For dual polarization and single channel transmission over $M$ lumped amplifier spans, the expression 
\begin{equation}
\eta = \frac{3\gamma^2}{\alpha^2}M^{1+\epsilon} \tanh\left(\frac{\alpha}{4|\beta_2| R_\tnr{s}^2}\right)
\label{nli:beygi}
\end{equation}
was proposed in \cite{beygi13}, and a qualitatively similar formula can be obtained from the results in \cite{poggiolini12}. Here, $\alpha$ is the attenuation coefficient of the fiber and the coefficient $\epsilon $ is between 0 and 1 (see \cite{poggiolini12,beygi13}) depending on how well the nonlinear interference decorrelates between each amplifier span. For single polarization transmission, the coefficient $3$ in (\ref{nli:beygi}) should be replaced by $2$ \cite{beygi12}.

The benefits of the GN model is that it is very accurate for uncoded transmission with traditional modulation formats,\footnote{The model is not valid for exotic modulation formats such as satellite constellations \cite{agrell12ipc}.} as demonstrated in experiments and simulations \cite{carena10,vacondio12,carena12}, and that it is very simple to analyze. It is, however, not intended for nonstationary input sequences, i.e., sequences whose statistics vary with time, because the transmit power $P$ in \eqref{infinite.memory.IO} is defined as the (constant) power of a random variable that generates the i.i.d.~symbols $x_k$. In order to capture the behavior of a wider class of transmission schemes, the GN model can be modified to depend on a time-varying transmit power, which is the topic of the next section.

\subsection{The Finite-Memory GN Model}\label{Sec:Model.Finite}

As mentioned in \secref{Sec:Introduction} and \ref{Sec:Model.Infinite}, a finite-memory model is essential in order to model the channel output corresponding to time-varying input distributions. Therefore, we refine the GN model in \secref{Sec:Model.Infinite} to make it explicitly dependent on the channel memory $N$, in such a way that the model ``converges'' to the regular GN model as $N\rightarrow\infty$. Many such models can be formulated.
In this paper, we aim for simplicity rather than accuracy.

The proposed model assumes that the input--output relation is still given by \eqref{eq:awgn}, but the average transmit power~$\Px$ in~\eqref{infinite.memory.IO} is replaced by an \emph{empirical power}, \ie by the arithmetic average of the squared magnitude of the symbol $x_{k}$ and of the $2N$ symbols around it. Mathematically,~\eqref{infinite.memory.IO} is replaced by
\begin{align}
\label{finite.memory.IO}
\Zn & =  \Znt\sqrt{\PASE + \eta  \left(\frac{1}{2N+1} \sum_{i=k-N}^{k+N} |x_i|^2\right)^{3}}
\end{align}
for any $k\in\mathbb{Z}$, where $N$ is the (one-sided) \emph{channel memory}. 
We refer to \eqref{eq:awgn} and \eqref{finite.memory.IO} as the \emph{finite-memory GN model}. %
Since (second-order) group velocity dispersion causes symmetric broadening with respect to the transit time of the signal, inter-symbol interference from dispersion will act both backwards and forwards in terms of the symbol index.
This is why both past and future inputs contribute to the noise power in~\eqref{finite.memory.IO}.
A somewhat related model for the additive noise in the context of data transmission in electronic circuits has been recently proposed in~\cite{koch09-08a}, where the memory is single-sided and the noise scales linearly with the signal power, not cubically as in \eqref{finite.memory.IO}.

Having introduced the finite-memory GN model, we now discuss some particular cases. First, the memoryless AWGN channel model can be obtained from both the GN and finite-memory GN models by setting $\eta=0$. 
In this case, the noise variance is $\Ex[|\Zn|^2]=\PASE$ for all $k$. 
Second, let us consider the scenario where the transmitted symbols is the random process $\{X_i\}$. 
Then the empirical power $(1/(2N+1))\sum_{i=k-N}^{k+N} |X_i|^2$ at any discrete time $k$ is a random variable that depends on the magnitude of the $k$th symbol and the $2N$ symbols around it. In the limit $N\rightarrow\infty$, this empirical power converges to the ``statistical'' power $\Px$ in \eqref{P.statistical}, for any i.i.d.~process with power $\Px$, as mentioned in \secref{Sec:Model.Infinite}. 
This observation shows that the proposed finite-memory model in \eqref{finite.memory.IO} ``converges'' to the GN model in \eqref{infinite.memory.IO}, provided that the channel memory $N$ is sufficiently large and that the process consists of i.i.d.~symbols with zero mean and variance $\Px$.

 The purpose of the finite-memory model is to be able to predict the output of the channel when the transmitted symbols are not i.i.d.~This is the case for example when the transmitted symbols are a nonstationary process (as will be exemplified in \secref{Sec:pulses}) and also for coded sequences (which we discuss in \secref{Sec:Capacity}). An advantage of the finite-memory model, from a theoretic viewpoint, is that the input--output relation of the channel is modeled as a fixed conditional probability of the output given the input and its history, which is the common notion of a channel model in communication and information theory ever since the work of Shannon \cite{shannon48}, \cite[p.~74]{gallager68}. This is in contrast to the regular GN model and other channel models, whose conditional distribution change depending on which transmitter the channel is connected to. Specifically, the GN model is represented by a family of such  conditional distributions, one for each value of the transmitter parameter $P$. %

\begin{figure}[tbp]
\newcommand{\scale}{0.85}
\psfrag{xlabel}[lB][ct][\scale][29]{Symbol slots}
\psfrag{ylabel}[rB][ct][\scale][-19]{Distance [km]}
\begin{center}
\includegraphics{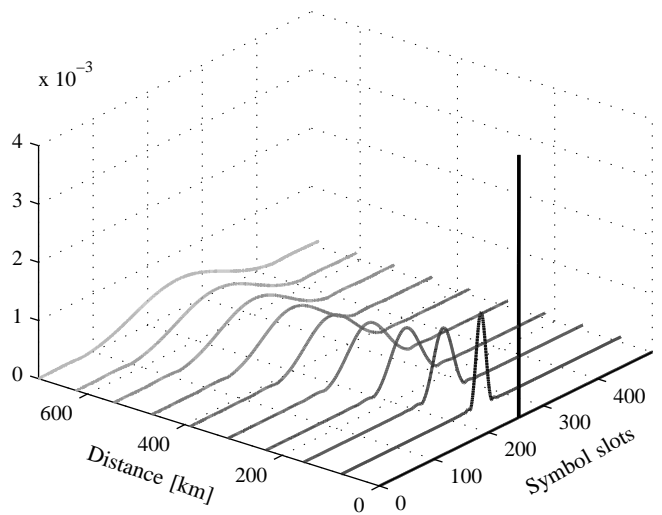}
\caption{Amplitude for a linearly propagating $15.6~\tnr{ps}$ raised-cosine pulse (compatible with $32~\tnr{GBaud}$) over $700~\tnr{km}$ fiber with $\beta_2 = -21.7~\tnr{ps}^2/\tnr{km}$. The lossy NLSE over 10 amplifier spans was simulated, with ASE noise switched off for clarity, and the peak power used was 0.1 mW.}
\label{pulse_broadening}
\end{center}
\end{figure}

A drawback with the proposed finite-memory model is that it is more complex than the GN model. Also, our model is not accurate for small values of $N$, since the GN assumption relies on the central limit theorem \cite{poggiolini11,beygi12,Johannisson13}. Furthermore, we assumed that all the $2N$ symbols around the symbol $x_{k}$ affect the noise variance equally. In practice, this is not the case. We nevertheless use the proposed model in this paper because it is relatively easy to analyze (see Sec.~\ref{Sec:Error} and \ref{Sec:Capacity}) and because even this simple finite-memory model captures the quantitative effects caused by non-i.i.d.~symbols, which is essential for the capacity analysis in \secref{Sec:Capacity}.

\subsection{Numerical Comparison} \label{Sec:pulses}

Before analyzing the finite-memory GN model, we first quantify the chromatic dispersion of the optical fiber. 
To this end, we simulated the transmission of a single symbol pulse over a over a single-channel, single-polarization fiber link without dispersion compensation. Ten amplifiers spans over a total distance of $700~\tnr{km}$ are simulated using the lossy NLSE model. We used a raised-cosine pulse with peak power $0.1~\tnr{mW}$ and a duration of $15.6~\tnr{ps}$ at half the maximum amplitude, which corresponds to half the symbol slot in a $32~\tnr{Gbaud}$ transmission system. The result is illustrated in \figref{pulse_broadening}. At this low power, the nonlinear effects are almost negligible. For clarity of illustration, the ASE noise was neglected by setting $\PASE=0$. The remaining system parameters are given in \tabref{tab:constants} and will be used throughout the paper, except when other values are explicitly stated. As we can see, the pulse broadens as it propagates along the fiber, having a width corresponding to about $100$ data symbols after $700~\tnr{km}$ of transmission, or a half-width of $N=50$ symbols. This is in good agreement with the relation for symbol memory used in \cite[p.~2037]{ip07}, which gives $2N \approx 2\pi |\beta_2| L R_\text{s}^2 = 97$.

\begin{table}
	\caption{System parameters used in the paper.}
	\centering
	\begin{tabular}{ c | c | l }
\hline
	Symbol	& Value & Meaning \\
\hline 

\hline
		$\alpha$			& $0.2~\tnr{dB/km}$ 			& Fiber attenuation \\
		$\beta_2$			& $-21.7~\tnr{ps}^2/\tnr{km}$ 	& Group velocity dispersion\\
		$\gamma$		& $1.27~\tnr{(W km)}^{-1}$ 	& Fiber nonlinear coefficient\\
		$M$				& $10$ 					& Number of amplifier spans\\
		$L$				& $700~\tnr{km}$ 			& System length\\
		$R_\tnr{s}$		& $32~\tnr{Gbaud}$			& Symbol rate\\
		$\PASE$ 			& $4.1\cd10^{-6}~\tnr{W}$		& Total ASE noise \\
		$\eta$	 		& $7244~\tnr{W}^{-2}$ 		& NLI coefficient \\
\hline

\hline
	\end{tabular}
	\label{tab:constants}
\end{table}

\begin{figure}[tbp]
\newcommand{\scale}{0.85}
\begin{center}
\psfrag{NLSE}[cl][cl][\scale]{\fcolorbox{black}{white}{{\color{black}NLSE simulation}}}
\psfrag{N=50}[cl][cl][\scale]{\fcolorbox{black}{white}{{\color{black}$N=50$}}}
\psfrag{GN}[cl][cl][\scale]{\fcolorbox{black}{white}{{\color{black}GN model}}}
\includegraphics[width=\columnwidth]{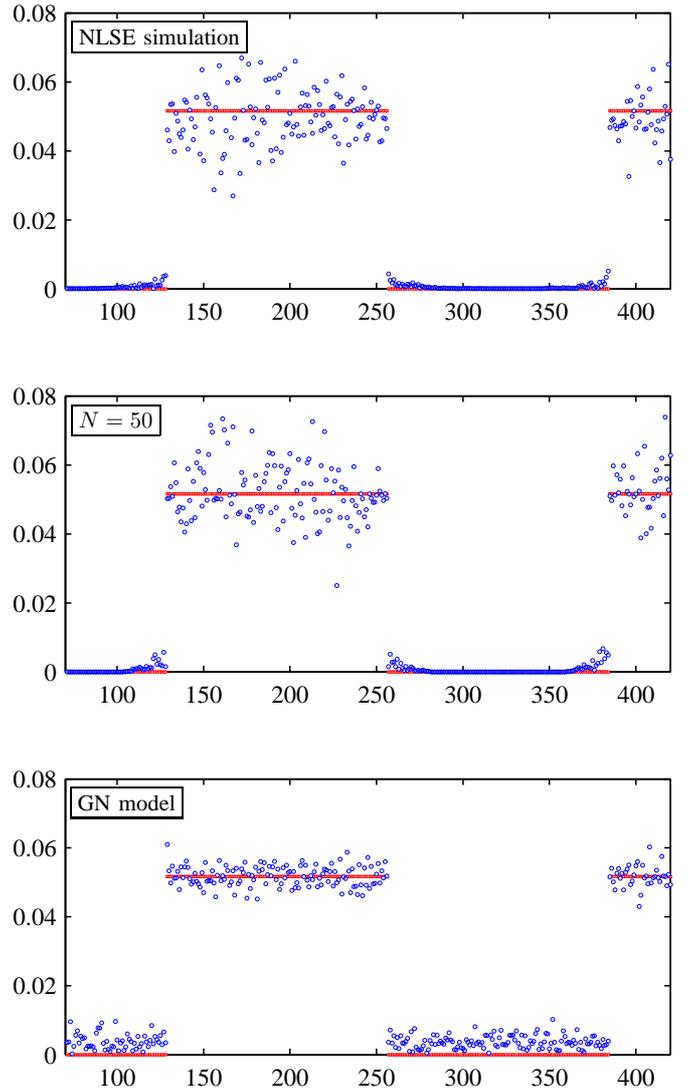}
\caption{Amplitude of the transmitted QPSK symbols $|x_{k}|$ (red squares) and received symbols $|\Yn|$ (blue circles) transmitted in a $700~\tnr{km}$ fiber link. The received symbols are obtained using (top) the NLSE, (middle) the finite-memory GN model \eqref{finite.memory.IO} with $N=50$, and (bottom) the regular GN model \eqref{infinite.memory.IO}.}
\label{pulsed_tx_example}
\end{center}
\end{figure}

Next, to validate the behavior of the finite-memory model with nonstationary input symbol sequences, we simulated the transmission of independent quadrature phase-shift keying (QPSK) data symbols with a time-varying magnitude, over the same $700$~km fiber link, at $R_\tnr{s}=32~\tnr{Gbaud}$. The transmitted sequence consists of $128$ symbols with $4~\tnr{mW}$ of average signal power, 128 symbols at $0~\tnr{mW}$ power, 128 symbols at $4~\tnr{mW}$, and so on. The statistical power is then $2~\tnr{mW}$. The chosen pulse shape is a raised-cosine return-to-zero pulse. 
In \figref{pulsed_tx_example}, we show the amplitude of the transmitted symbols $|x_{k}|$ (red) and received symbols $|\Yn|$ (blue) with three different models: the NLSE, the finite-memory GN model with $N=50$, and the regular GN model. In the middle and lower plots of \figref{pulsed_tx_example}, we used the NLI coefficient $\eta = 7244~\tnr{W}^{-2}$, which was calculated from \eqref{nli:beygi}, using $\epsilon=0$ for simplicity. Also in \figref{pulsed_tx_example}, we used $\PASE=0$ to better illustrate the properties of the nonlinear models.

As can be seen, the agreement between the NLSE simulations and the finite memory model is quite reasonable, but the GN model cannot capture the nonstationary dynamics.
The results in \figref{pulsed_tx_example} also show that the noise variance in the NLSE simulation is low around the symbols with low input power and high around the symbols with high input power. This behavior is captured by the finite-memory GN model but not by the regular GN model, for which the variance of the noise is the same for any time instant. This illustrates that the GN model~\eqref{infinite.memory.IO} should be avoided with nonstationary symbol sequences as the ones used in~\figref{pulsed_tx_example}. This is not surprising, as the model was derived under an i.i.d.\ assumption. In Sec.~\ref{Sec:Conclusions}, we will return to this observation when analyzing coded transmission. We believe that the finite-memory GN model proposed here, albeit idealized, is the first model that is able to deal with nonstationary symbol sequences.

\section{Uncoded Error Probability}\label{Sec:Error}

We assume that the transmitted symbols $\{\Xn\}$ are independently drawn from a discrete constellation $\mcS=\set{s_{1},\ld,s_{M}}$, where $M=2^{m}$. The symbols are assumed to be selected with the same probability, and thus, the average transmit (statistical) power is given by
\begin{align}\label{Px}
\Px = \Ex[|\X|^{2}]=\frac{1}{M}\sumi |s_{i}|^{2}.
\end{align}
For each time instant $k$, we denote the sequence of the $2N$ symbols transmitted around $x_{k}$ by 
\begin{align}\label{x.mem.seq}
\Xmem \triangleq [\XkmN,\ld,\Xkmo,\Xkpo,\ld,\XkpN],
\end{align}
where the notation emphasizes that $\Xmem$ is a random vector describing the channel memory at time instant $k$. 
For future use, we define the function
\begin{align}
\label{rho.func.def}
{\rho}(a)	& \triangleq \PASE+ \eta\left(\frac{a}{2N+1}\right)^{3}.
\end{align}

For a given sequence of $2N$ symbols $\xmem$ and a given transmitted symbol $X_k=s_{i}$, the conditional variance of the additive noise in \eqref{finite.memory.IO} can be expressed as an explicit function of $\xmem$ using \eqref{rho.func.def}, \ie
\begin{align}
\label{noise.variance}
\Szsilower	& = \PASE+ \eta\left(\frac{|s_{i}|^{2}\!+\|\xmem\|^{2}}{2N+1}\right)^{3},
\end{align}
where $\|\bx\|$ denotes the Euclidean norm of $\bx$.
For a given transmitted symbol $X_k=s_{i}$ and a given sequence $\xmem$, the channel law for the finite-memory model is
\eqlab{ctp}{
\nonumber
&\pdf_{\Yn|\Xn,\Xmem}(\y|s_{i},\xmem) \\
&\qquad	\triangleq \frac{1}{\pi\Szsilower}\exp{\biggl(-\frac{|\y-s_{i}|^{2}}{\Szsilower}\biggr)}.
}

\subsection{Error Probability Analysis}\label{Sec:Error.BEP.and.SEP}

\begin{figure}[tbp]
\newcommand{\scale}{0.85}
\psfrag{dd}[cb][cb][\scale]{$2\Delta$}
\psfrag{x1}[cb][cb][\scale]{$s_1$}\psfrag{x2}[cb][cb][\scale]{$s_2$}
\psfrag{x3}[cb][cb][\scale]{$s_3$}\psfrag{x4}[cb][cb][\scale]{$s_4$}
\psfrag{x5}[cb][cb][\scale]{$s_5$}\psfrag{x6}[cb][cb][\scale]{$s_6$}
\psfrag{x7}[cb][cb][\scale]{$s_7$}\psfrag{x8}[cb][cb][\scale]{$s_8$}
\psfrag{x9}[cb][cb][\scale]{$s_9$}\psfrag{x10}[cb][cb][\scale]{$s_{10}$}
\psfrag{x11}[cb][cb][\scale]{$s_{11}$}\psfrag{x12}[cb][cb][\scale]{$s_{12}$}
\psfrag{x13}[cb][cb][\scale]{$s_{13}$}\psfrag{x14}[cb][cb][\scale]{$s_{14}$}
\psfrag{x15}[cb][cb][\scale]{$s_{15}$}\psfrag{x16}[cb][cb][\scale]{$s_{16}$}
\psfrag{b1}[cb][cb][\scale]{{\color{red}00}{\color{blue}00}}
\psfrag{b2}[cb][cb][\scale]{{\color{red}00}{\color{blue}01}}
\psfrag{b3}[cb][cb][\scale]{{\color{red}00}{\color{blue}11}}
\psfrag{b4}[cb][cb][\scale]{{\color{red}00}{\color{blue}10}}
\psfrag{b5}[cb][cb][\scale]{{\color{red}01}{\color{blue}00}}
\psfrag{b6}[cb][cb][\scale]{{\color{red}01}{\color{blue}01}}
\psfrag{b7}[cb][cb][\scale]{{\color{red}01}{\color{blue}11}}
\psfrag{b8}[cb][cb][\scale]{{\color{red}01}{\color{blue}10}}
\psfrag{b9}[cb][cb][\scale]{{\color{red}11}{\color{blue}00}}
\psfrag{b10}[cb][cb][\scale]{{\color{red}11}{\color{blue}01}}
\psfrag{b11}[cb][cb][\scale]{{\color{red}11}{\color{blue}11}}
\psfrag{b12}[cb][cb][\scale]{{\color{red}11}{\color{blue}10}}
\psfrag{b13}[cb][cb][\scale]{{\color{red}10}{\color{blue}00}}
\psfrag{b14}[cb][cb][\scale]{{\color{red}10}{\color{blue}01}}
\psfrag{b15}[cb][cb][\scale]{{\color{red}10}{\color{blue}11}}
\psfrag{b16}[cb][cb][\scale]{{\color{red}10}{\color{blue}10}}
\begin{center}
\includegraphics{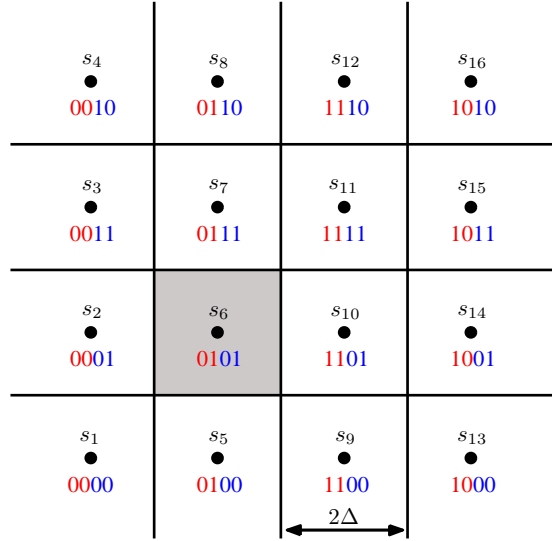}
\caption{The 16-QAM constellation $\mcS$ and its binary labeling. The binary labeling of the constellation is based on the Cartesian product of the BRGC for 4-ary pulse amplitude modulation in phase (red) and quadrature (blue). The Voronoi regions of the symbols and the MED of the constellation are also shown. The Voronoi region $\mcV_{6}$ is highlighted in gray. %
}
\label{16QAM}
\end{center}
\end{figure}

We consider the equally spaced 16-QAM constellation shown in \figref{16QAM}. In this case, $\mcS=\set{a+b\sqrt{-1} : a,b \in \set{\pm\Delta,\pm3\Delta}}$, the minimum Euclidean distance (MED) of the constellation is $2\Delta$, and the statistical power is $\Px=10\Delta^{2}$. The binary labeling is the binary reflected Gray code (BRGC) \cite{Agrell04}, where the first two bits determine the in-phase (real) component of the symbols and the last two bits determine the quadrature (imaginary) components of the symbols. This is shown with colors in \figref{16QAM}.

The maximum-likelihood (ML) symbol-by-symbol detection rule for a given sequence $\xmem$ chooses the symbol $s_{i}\in\mcS$ that maximizes $\pdf_{\Yn|\Xn,\Xmem}(\y|s_{i},\xmem)$ in \eqref{ctp}. The decision made by this detector can be expressed as
\begin{IEEEeqnarray}{rCL}\label{ml.rule}
 \hX_{k}^{\tr{ML}} &=& \argmin_{s_{i}\in\mcS}\Biggl\{\log{\Szsilower} \notag\\
 &&+ \frac{|\y-s_{i}|^{2}}{\Szsilower}\Biggr\}, 
\end{IEEEeqnarray}
which shows that, due to the dependency of $\log{\Szsilower}$ on $s_{i}$, this detector is not an MED detector. For simplicity, however, we disregard this term and study the MED detector, which chooses the symbol $s_{i}$ being closest, in Euclidean distance, to the channel output $y$. Thus
\begin{align}\label{med.Voronoi}
\hX_{k} &= \argmin_{s_i \in \mcS} |y-s_i|^2 \notag\\
&=s_{i}, \quad \text{if $\Y_{k}\in\mcV_{i}$},
\end{align}
where $\mcV_i$ denotes the decision region, or \emph{Voronoi region,} of $s_i$.

\begin{remark}
As we will later see, for  memory $N$, the MED detector in \eqref{med.Voronoi} is in fact equivalent to the detector in \eqref{ml.rule}. Intuitively, this holds because the approximation $\|\xmem\|^{2}+|s_{i}|^{2}\approx\|\xmem\|^{2}$ becomes tight when $N$ is large.
\end{remark}

\begin{remark}
The ML symbol-by-symbol detector in \eqref{ml.rule} is suboptimal, i.e., better detectors can be devised. For example, one could design a detector that uses not only the current received symbol, but also the next $N$ received symbols. Since the current transmitted symbol will affect the noise of the next $N$ symbols, this information could be taken into account to make a better decision on the current symbol. In this paper, however, we focus on the MED detector in \eqref{med.Voronoi} because of its simplicity.
\end{remark}

The following two theorems give closed-form expressions for the BER and SER for the constellation in \figref{16QAM} when used over the finite-memory GN model.
\begin{theorem}\label{BEP.16QAM.theorem}
For the finite-memory GN model with arbitrary memory $N<\infty$, the BER of the MED detector for the 16-QAM constellation in \figref{16QAM} is given by
\begin{align}
& \BEP = \frac{2^{-3}}{2^{4N}}\suml \nchoosek{4N}{l} \sum_{\substack{r\in\set{1,3,5}\\t\in\set{1,5,9}}}B_{r,t}\QF\lefto(\sqrt{\frac{r^{2}\Px}{5\gammaltN{l}{t}{N}}}\right),
\label{BEP.16QAM}
\end{align}
where 
\begin{align}
\label{B.1}
B_{1,1}&=2,~B_{3,1}=1,~B_{5,1}=0,\\
\label{B.5}
B_{1,5}&=3,~B_{3,5}=2,~B_{5,5}=-1,\\
\label{B.9}
B_{1,9}&=1,~B_{3,9}=1,~B_{5,9}=-1,
\end{align}
and where 
\begin{align}\label{gamma}
\gammaXNl \triangleq \PASE+ \frac{\eta}{(2N+1)^{3}} \biggl(\frac{\Px(2N+4l+t)}{5}\biggr)^{3}.
\end{align}
\end{theorem}
\begin{IEEEproof}
See \appref{BEP.16QAM.theorem.Proof}.
\end{IEEEproof}

\begin{theorem}\label{SEP.16QAM.theorem}
For the finite-memory GN model with arbitrary memory $N<\infty$, the SER of the MED detector for the 16-QAM constellation in \figref{16QAM} is given by
\begin{align}
\SEP &= \frac{4^{-1}}{4^{2N}}\suml \nchoosek{4N}{l} \sum_{\substack{e\in\set{1,2}\\t\in\set{1,5,9}}}S_{e,t}\QF\lefto(\sqrt{\frac{\Px}{5\gammaltN{l}{t}{N}}}\right)^{e}
\label{SEP.16QAM},
\end{align}
where 
\begin{align}
\label{S.1}
S_{1,1}&=4,~B_{2,1}=-4,\\
\label{S.5}
S_{1,5}&=6,~B_{2,5}=-4,\\
\label{S.9}
S_{1,9}&=2,~B_{2,9}=-1,
\end{align}
and where $\gammaXNl$ is given by \eqref{gamma}.
\end{theorem}
\begin{IEEEproof}
See \appref{SEP.16QAM.theorem.Proof}.
\end{IEEEproof}

The BER and SER in the limit $N\to\infty$ can be inferred from \theosref{BEP.16QAM.theorem}{SEP.16QAM.theorem} as shown in the next corollary.
\begin{corollary}\label{16QAM.infmemory.theorem}
The BER and SER for the finite-memory GN model in the limit $N\to\infty$ are
\begin{align}
  \BEP &=
  \frac{3}{4}\QF\lefto(\sqrt{\frac{\Px/5}{\PASE + \eta \Px^3}}\right)+\frac{1}{2}\QF\lefto(\sqrt{\frac{9\Px/5}{\PASE + \eta \Px^3}}\right) \notag\\
  \label{BEP.16QAM.infmemory}
  & \qquad -\frac{1}{4}\QF\lefto(\sqrt{\frac{5\Px}{\PASE + \eta \Px^3}}\right),\\
  \label{SEP.16QAM.infmemory}
  \SEP &=
  3\QF\lefto(\sqrt{\frac{\Px/5}{\PASE + \eta \Px^3}}\right)-\frac{9}{4}\QF\lefto(\sqrt{\frac{\Px/5}{\PASE + \eta \Px^3}}\right)^{2}.
\end{align}
\end{corollary}
\begin{IEEEproof}
See \appref{16QAM.infmemory.theorem.Proof}.
\end{IEEEproof}

The other extreme case to consider is the memoryless AWGN channel. The BER and SER expressions in this case are given in the following corollary.
\begin{corollary}\label{16QAM.nomemory.theorem}
The BER and SER for the memoryless AWGN channel are given by
\begin{align}
\nonumber
\BEP &= \frac{3}{4}\QF\lefto(\sqrt{\frac{\Px}{5\PASE}}\right)+\frac{1}{2}\QF\lefto(\sqrt{\frac{9\Px}{5\PASE}}\right)\\
\label{BEP.16QAM.nomemory}
  & \qquad -\frac{1}{4}\QF\lefto(\sqrt{\frac{5\Px}{\PASE}}\right),\\
\label{SEP.16QAM.nomemory}
\SEP &= 3\QF\lefto(\sqrt{\frac{\Px}{5\PASE}}\right)-\frac{9}{4}\QF\lefto(\sqrt{\frac{\Px}{5\PASE}}\right)^{2}.
\end{align}
\end{corollary}
\begin{IEEEproof}
Set $\eta=0$ in~\eqref{BEP.16QAM.infmemory} and~\eqref{SEP.16QAM.infmemory}.
\end{IEEEproof}

The results in Corollaries~\ref{16QAM.infmemory.theorem} and \ref{16QAM.nomemory.theorem} correspond to the well-known expressions for the BER and SER for the AWGN channel.
In particular, 
\eqref{BEP.16QAM.nomemory} can be found in \cite[eq.~(10)]{Fitz94}, \cite[eq.~(10.36a)]{Simon95_Book} and \eqref{SEP.16QAM.nomemory} in \cite[eq.~(10.32)]{Simon95_Book}. 
Also, the results in Corollary~\ref{16QAM.nomemory.theorem} together with \eqref{infinite.memory.IO} show that the BER and SER for the finite-memory GN model when $N\rightarrow\infty$ converge to the BER and SER for the regular GN model.

\begin{figure}[tbp]
\begin{center}
\newcommand{\scale}{0.85}
\psfrag{xlabel}[cc][cB][\scale]{$\Px$~[dBm]}%
\psfrag{ylabel}[cc][cB][\scale]{$\BEP$}
\psfrag{N=0}[cl][cl][\scale]{AWGN}%
\psfrag{N=1}[cl][cl][\scale]{$N=1$}%
\psfrag{N=2}[cl][cl][\scale]{$N=2$}%
\psfrag{N=5}[cl][cl][\scale]{$N=5$}%
\psfrag{N=10}[cl][cl][\scale]{$N=10$}%
\psfrag{N=50}[cl][cl][\scale]{$N=50$}%
\psfrag{N=infinityyyy}[cl][cl][\scale]{GN model}%
\includegraphics{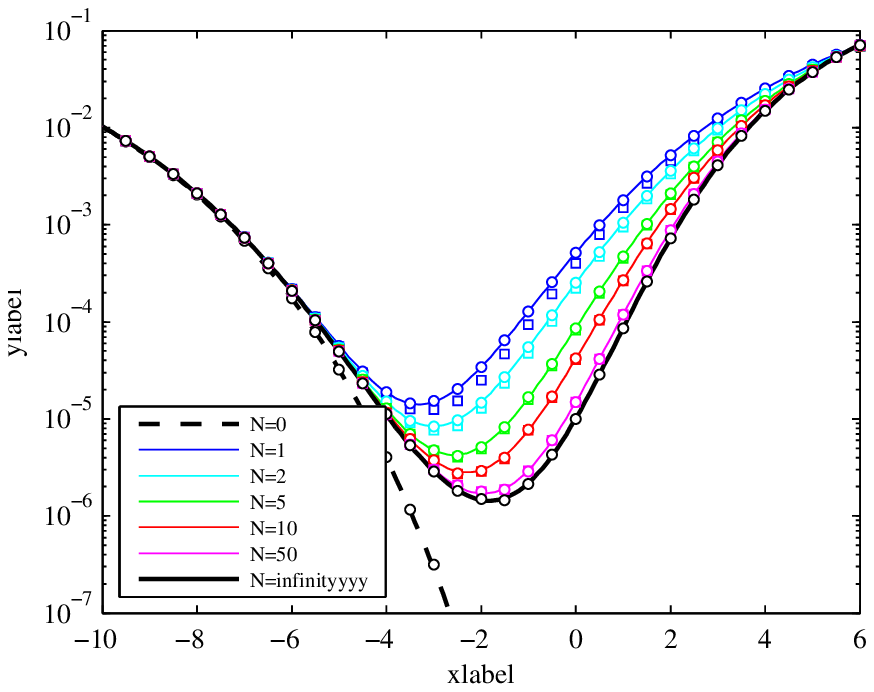}
\psfrag{ylabel}[cc][cB][\scale]{$\SEP$}
\includegraphics{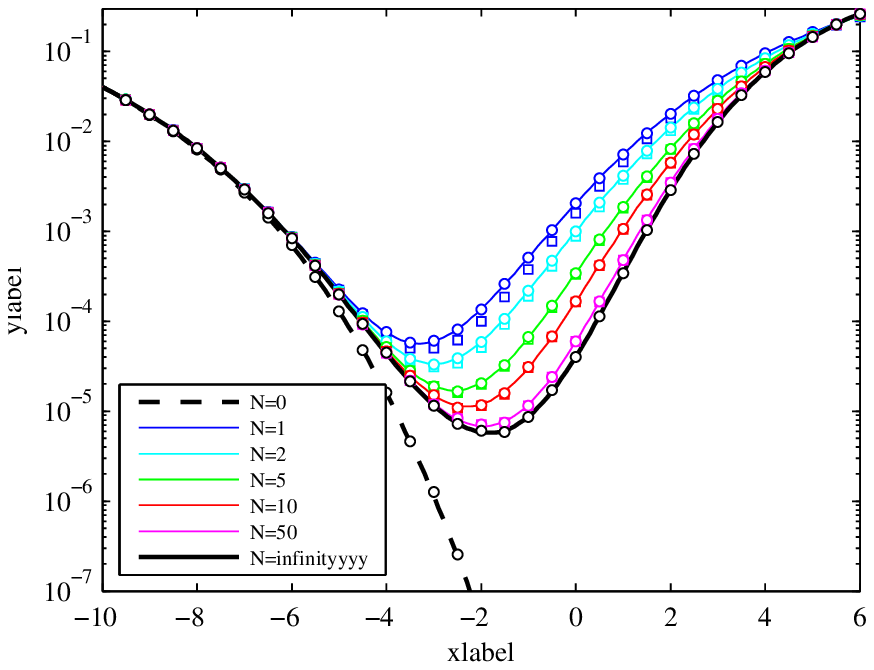}
\caption{Analytical BER (top) and SER (bottom) of 16-QAM transmission with the finite-memory GN model, for different values of $N$ (solid lines). Markers show simulation results with the ML detector in \eqref{ml.rule} (squares) and the MED detector in \eqref{med.Voronoi} (circles). The results for the memoryless AWGN channel and the regular GN model are included for comparison.}
\label{BER.SEP.16QAM}
\end{center}
\end{figure}

\subsection{Numerical Results}\label{Sec:Error.Results}

We consider the same scenario as in \secref{Sec:pulses}, with parameters according to \tabref{tab:constants}.
The BER and SER for the 16-QAM constellation in \figref{16QAM} given by \theosref{BEP.16QAM.theorem}{SEP.16QAM.theorem} are shown in \figref{BER.SEP.16QAM} for different values of $N$. \figref{BER.SEP.16QAM} also shows the asymptotic cases $N=0$ and $N\rightarrow\infty$ given by Corollaries~\ref{16QAM.infmemory.theorem} and~\ref{16QAM.nomemory.theorem}, respectively. 
Furthermore, results obtained via computer simulations of \eqref{eq:awgn}--\eqref{infinite.memory.IO} are included using the ML detector in \eqref{ml.rule}, marked with squares, and the MED detector in \eqref{med.Voronoi}, marked with circles. As expected, the MED detector yields a perfect match with the analytical expressions, whereas the ML detector deviates slightly for small channel memories.
 
The results in \figref{BER.SEP.16QAM} show that in the low-input-power regime, the memory in the channel plays no role for the BER and SER, and all the curves follow closely the BER and the SER of a memoryless AWGN channel. However, as $\Px$ increases, the memory kicks in, causing the BER and SER for finite $N$ to have a minimum, and then to increase as $\Px$ increases. 
Physically, this can be explained as follows: in the low-power regime, the BER is limited by the ASE noise, which is independent of the memory depth. In the high-power regime, the Kerr-induced noise dominates, resulting in increasing BER with power. Similar behavior has been reported in most experiments and simulations on nonlinearly-limited links, e.g., \cite{mecozzi94, demir07, carena12, beygi12}, \cite[Ch.~9]{agrawal10}.
The reason why the performance improves slightly with the memory depth $N$ is the nonlinear scaling of the Kerr-induced noise. For $N=1$, sequences of two or more high-amplitude symbols will receive high noise power and dominate the average BER. For higher $N$, longer (and less probable) sequences of high-amplitude symbols are required to receive the same, high, noise power. Thus on average the performance improves with $N$, up to a limit given by the GN model.

The results in \figref{BER.SEP.16QAM} also show how the finite-memory model in the high-input power regime approaches the GN model. For $N=50$, the two models yield very similar BER and SER curves.

\section{Channel Capacity}\label{Sec:Capacity}

In this section, some fundamentals of information theory are first reviewed. Then a lower bound on the capacity of the finite-memory GN model is derived and evaluated numerically.

\subsection{Preliminaries}\label{Sec:Capacity:Prel}
\begin{figure}[tbp]
\newcommand{\scale}{0.85}
\psfrag{ENC}[cc][cc][\scale]{Encoder}
\psfrag{DEC}[cc][cc][\scale]{Decoder}
\psfrag{COMM}[cc][cc][\scale]{Physical}
\psfrag{CHANNEL}[cc][cc][\scale]{Channel}
\psfrag{x}[bc][bc][\scale]{$\bx$}
\psfrag{y}[bc][Bc][\scale]{$\bY$}
\psfrag{z}[bl][Bl][\scale]{$\bZ$}
\psfrag{j}[bc][Bc][\scale]{$j$}
\psfrag{jh}[bc][Bc][\scale]{$\hat{\jmath}$}
\begin{center}
\includegraphics[width=\columnwidth]{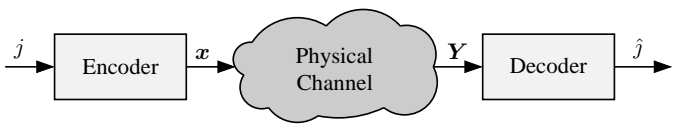}
\caption{Encoder and decoder pair. The encoder maps a message $j$ to a codeword $\bx=[x_1,\dots,x_n]$. The decoder uses the noisy observation $\bY=[Y_1,\dots,Y_n]$ to provide an estimate $\hat{\jmath}$ of the message $j$ .}
\label{enc_dec}
\end{center}
\end{figure}

\figref{enc_dec} shows a generic coded communication system where a message $j$ is mapped to a codeword $\bx=[x_1,\dots,x_n]$. 
This codeword is then used to modulate  a continuous-time waveform, which is then transmitted through the physical channel. 
At the receiver's side, the continuous-time waveform is processed (filtered, equalized, synchronized, matched filtered, sampled, etc.) resulting in a discrete-time observation $\bY=[Y_1,\dots,Y_n]$, which is a noisy version of the transmitted codeword $\bx$. 
The decoder uses $\bY$ to estimate the transmitted message $j$.

 When designing a coded communication system, the first step is to choose the set of codewords (i.e., the \emph{codebook}) that will be transmitted through the channel. 
 Once the codebook has been chosen, the mapping rule between messages and codewords should be chosen, which fully determines the encoding procedure. 
At the receiver side, the decoder block will use the mapping rule used at the transmitter (as well as the channel characteristics) to give an estimate~$\hat{\jmath}$ of the message~$j$. 
The triplet codebook, encoder, and decoder forms a so-called \emph{coding scheme}.
Practical coding schemes are designed so as to minimize the probability that $\hat{\jmath}$ differs from $j$, while at the same time keeping the complexity of both encoder and decoder low.

Channel capacity is the largest transmission rate at which reliable communications can occur. More formally, let $(n,M,\epsilon)$ be a coding scheme consisting of:
\begin{itemize}
\item An encoder that maps a message $j\in\{1,\dots,M\}$ into a block of $n$ transmitted symbols $\bx = [x_1,\dots,x_n]$ satisfying a per-codeword power constraint
\begin{IEEEeqnarray}{rCL}\label{eq:avp}
\frac{1}{n}\sum_{l=1}^{n}|x_l|^2=\Px.
\end{IEEEeqnarray}
  \item A decoder that maps the corresponding block of received symbols $\bY=[Y_1,\dots,Y_n]$ into a message $\hat{\jmath}\in\{1,\dots,M\}$ so that the average error probability, \ie the probability that $\hat{\jmath}$ differs from $j$, does not exceed $\epsilon$.
\end{itemize}
Observe that $P$ here is defined differently from in previous sections. It still represents the average transmit power, but while this quantity is \secref{Sec:Model}--\ref{Sec:Error} was interpreted in a statistical sense as the mean of an i.i.d.~random variable, it is in this section the exact power of \emph{every} codeword.

The maximum coding rate $R^*(n,\epsilon)$ (measured in bit/symbol) for a given block length $n$ and error probability $\epsilon$ is defined as the largest ratio $(\log_2 M)/n$ for which an $(n,M,\epsilon)$ coding scheme exists.
The channel capacity $C$ is the largest coding rate for which a coding scheme with vanishing error probability exists, in the limit of large block length,
\begin{IEEEeqnarray}{rCL}\label{C.def}
C \triangleq \lim_{\epsilon\to 0}\lim_{n\to\infty}R^*(n,\epsilon).
\end{IEEEeqnarray}

\subsection{Memoryless Channels}\label{Sec:Capacity:NoMem}

By Shannon's \emph{channel coding theorem,} the channel capacity of a discrete-time memoryless channel, in bit/symbol, can be calculated as \cite{shannon48}, \cite[Ch.~7]{cover06}
\eqlab{eq:c}{
C = \sup I(X;Y)
,}
where $I(X;Y)$ is the \emph{mutual information} (MI)
\eqlab{eq:ixy}{
I(X;Y) = \iint f_{X,Y}(x,y) \log_2\frac{f_{X,Y}(x,y)}{f_X(x)f_Y(y)} dx dy
}
and the maximization in~\eqref{eq:c} is over all probability distributions $f_X$ that satisfy $\E[|X|^2]=\Px$, for a given channel $f_{Y|X}$.

Roughly speaking, a transmission scheme that operates at an arbitrary rate $R < C$ can be designed by creating a codebook of $M = 2^{nR}$ codewords of length $n$, whose elements are i.i.d.\ random samples from the distribution $f_X$ that maximizes the mutual information in \eqref{eq:c}. This codebook is stored in both the encoder and decoder. During transmission, the encoder maps each message $j$ into a unique codeword $\bx$, and the decoder identifies the codeword that is most similar, in some sense, to the received vector $\bY$. An arbitrarily small error probability $\epsilon$ can be achieved by choosing $n$ large enough. This \emph{random coding} paradigm was proposed already by Shannon \cite{shannon48}. In practice, however, randomly constructed codebooks are usually avoided for complexity reasons.

Since the additive noise in~\eqref{infinite.memory.IO} is statistically independent of $\Xn$, the channel capacity of the GN model \eqref{infinite.memory.IO} can be calculated exactly as \cite{splett93,bosco11}
\eqlab{eq:cawgn}{
C = \log_2\lefto(1+\frac{\Px}{\PASE + \eta \Px^3}\right)
}
using Shannon's well-known capacity expression \cite[Sec.~24]{shannon48}, \cite[Ch.~9]{cover06}. The capacity in \eqref{eq:cawgn} can be achieved by choosing the codewords $\bx$ to be drawn independently from a Gaussian distribution $\mc{CN}(0,\Px)$.

Considered as a function of the transmitted signal power~$\Px$, the capacity in \eqref{eq:cawgn} has the peculiar behavior of reaching a peak and eventually decreasing to zero at high enough power, since the denominator of \eqref{eq:cawgn} increases faster than the numerator. This phenomenon, sometimes called the ``nonlinear Shannon limit'' in the optical communications community, conveys the message that reliable communication over nonlinear optical channels becomes impossible at high powers.
In the following sections, we shall question this pessimistic conclusion.

\subsection{Channels with Memory}\label{Sec:Capacity:Mem}

The capacity of channels with memory is, under certain assumptions on information stability \cite[Sec. I]{verdu94},
\eqlab{eq:cmemory}{
C = \lim_{n\rightarrow\infty} \sup \frac{1}{n} I(\bX_1^n;\bY_1^n)
,}
where $\bX_i^j = (X_i,X_{i+1},\ldots,X_j)$, $I(\bX_i^j;\bY_i^j)$ is defined as a multidimensional integral analogous to \eqref{eq:ixy}, and the maximization is over all joint distributions of $X_1,\ldots,X_{n}$ satisfying~$\Ex\lefto[\|\bX_1^n\|^2\right]=n\Px$. In this context, it is worth emphasizing that the maximization in \eqref{eq:cmemory} includes sequences $X_1,\ldots,X_{n}$ that are not i.i.d. Hence, in order to calculate the channel capacity of a transmission link, it is essential that the employed channel model allows non-i.i.d.\ inputs.

An exact expression for the channel capacity of the finite-memory GN model \eqref{finite.memory.IO} is  not available. Shannon's formula, which leads to~\eqref{eq:cawgn}, does not apply here, because the sequences $\{\Xn\}$ and $\{\Zn\}$, where $\Zn$ was defined in~\eqref{finite.memory.IO}, are dependent. A capacity estimation via \eqref{eq:cmemory} is numerically infeasible, since it  involves integration and maximization over high-dimensional spaces. We therefore turn our attention to bounds on the capacity for the finite-memory model. Every joint distribution of $X_1,\ldots,\Xn$ satisfying~$\Ex\lefto[\|\bX_1^n\|^2\right]=n\Px$ gives us a lower bound on capacity. Thus,
\eqlab{eq:lbmemory}{
C \ge \lim_{n\rightarrow\infty} \frac{1}{n} I(\bX_1^n;\bY_1^n)
,}
for any random process $\{X_k\}$ such that the limit exists.

\begin{figure}
\begin{center}
\psfrag{Re}[][][.7][25]{$\mathrm{Re}\{X_k\}$}
\psfrag{Im}[][][.7][25]{$\mathrm{Im}\{X_k\}$}
\psfrag{k}[][][.7]{$k$}
{\includegraphics[width=\columnwidth]{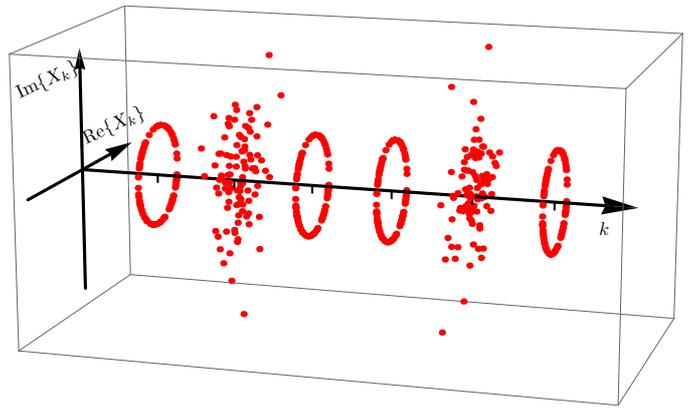}} %
\caption{Six samples of the random input process $\{X_k\}$ used to generate the lower bound in \theoref{th:clb}. The channel memory is here $N=1$, meaning that $2N+1=3$ input symbols $X_k$ influence each output symbol. The distributions are illustrated as scatter plots of $1000$ realizations for each sample.}
\label{fig:random-sequence}
\end{center}
\end{figure}

\subsection{Lower Bound}\label{Sec:Capacity:Bounds}
In this section, a lower bound on~\eqref{eq:cmemory} is derived by applying \eqref{eq:lbmemory} to the following random input process. In every block of $2N+1$ consecutive symbols, we let the first $N$ symbols and the last $N$ symbols have a constant amplitude, whereas the amplitude of the symbol in the middle of the block follows an arbitrary distribution.
The phase of each symbol in the block is assumed uniform.
With this random input process, illustrated in \figref{fig:random-sequence}, the memory in~\eqref{finite.memory.IO} depends only on a single variable-amplitude symbol. 
This enables us to derive an analytical expression for the resulting capacity lower bound in \eqref{eq:lbmemory}.
\begin{theorem}\label{th:clb}
For every $r_1\ge0$ and every probability distribution $f_R$ over $\reals^+$ such that
\begin{IEEEeqnarray}{rCL}\label{eq:power_normalization}
  \frac{2N r_1^2+\E[R^2]}{2N+1}=P,
\end{IEEEeqnarray}
where $R\sim f_R$, the channel capacity of \eqref{finite.memory.IO} is lower-bounded as
\begin{multline}\label{eq:clb}
C \ge -\frac{\E[\log_2 f_{\bU}(\bU)]}{2N+1} \\- \int_0^\infty f_R(r) \log_2 (e \rho(2Nr^2_{1}+r^2)) \,\tr{d}r.
\end{multline}
Here, $\bU\triangleq [U_{-N},U_{-N+1},\ldots,U_{N}]$ is a random vector whose probability density function $f_{\bU}$ is
\begin{IEEEeqnarray}{rCL}\label{fy.th.clb}
f_{\bU}(\bu)=&&\int_0^{\infty}f_R(r)\frac{\exp\lefto(-\frac{\sum_{k=-N}^N u_k+2Nr_1^2+r^2 }{\rho(2Nr^2_1+r^2)}\right)}{\bigl(\rho(2Nr^2_1+r^2)\bigr)^{2N+1}}\notag\\
  &&\cdot I_0\lefto( \frac{2r\sqrt{u_0}}{\rho(2N r_1^2+r^2)}\right) \notag\\
  &&\cdot \prod_{\substack{k=-N\\k\ne0}}^{N} I_0\lefto( \frac{2r_1 \sqrt{u_k}}{\rho(2Nr_1^2+r^2)}\right) \tr{d}r,
\end{IEEEeqnarray}
where the function $\rho(\cdot)$ is defined in \eqref{rho.func.def}, and $I_0(u)$ is the modified Bessel function of the first kind.
\end{theorem}

\begin{IEEEproof}
See Appendix \ref{app:clb}.
\end{IEEEproof}

The bound will be numerically computed in the next section.

\subsection{Numerical Results}\label{Sec:Capacity:Res}
Theorem~\ref{th:clb} yields a lower bound on capacity for every constant $r_1$ and every probability distribution $f_R$ satisfying~\eqref{eq:power_normalization}. Instead of optimizing the bound over all distributions $f_R$, which is of limited interest, since the theorem itself provides only a lower bound on capacity, we study a heuristically chosen family of distributions and optimize its parameters along with the constant amplitude $r_1$.

\begin{figure}
\begin{center}
\psfrag{MI}[][][.7]{$C$}
\psfrag{v}[][][.7][10]{$\nu$}
\psfrag{rs}[l][b][.7][-60]{$r_1^2/s$}
\begin{tabular}{cc}
\includegraphics[width=4cm]{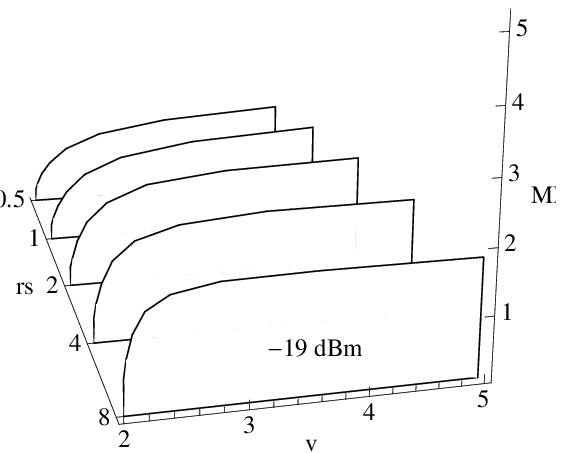} & \includegraphics[width=4cm]{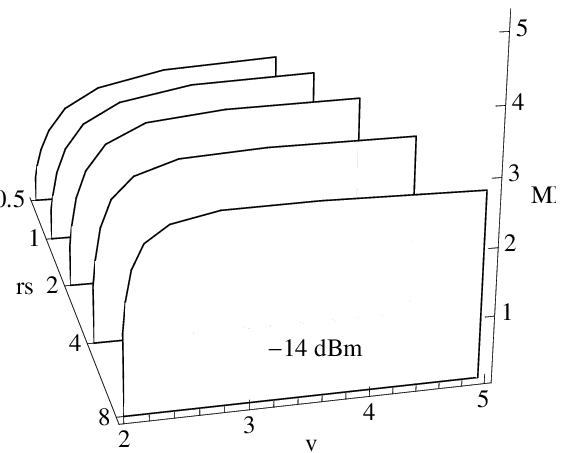} \\
\includegraphics[width=4cm]{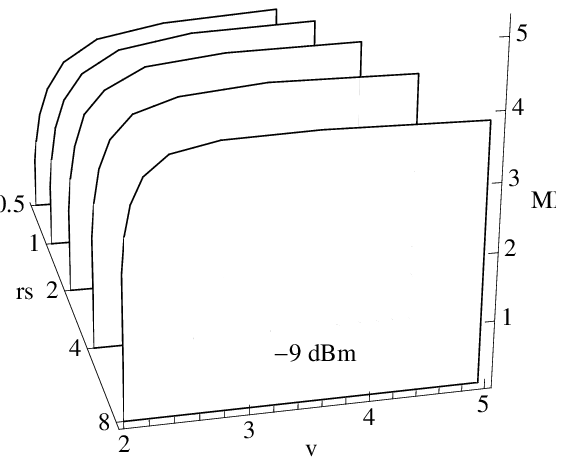} & \includegraphics[width=4cm]{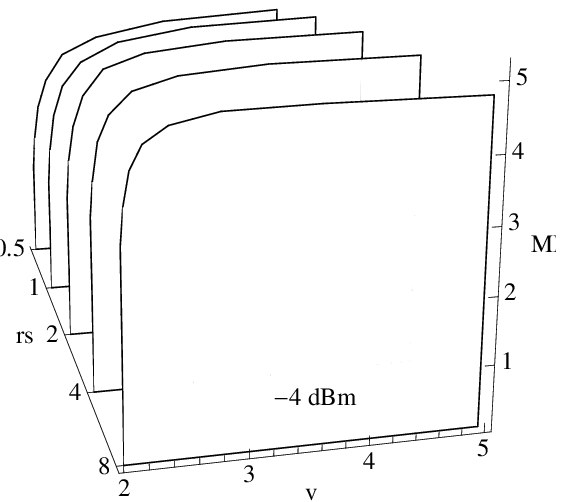} \\
\includegraphics[width=4cm]{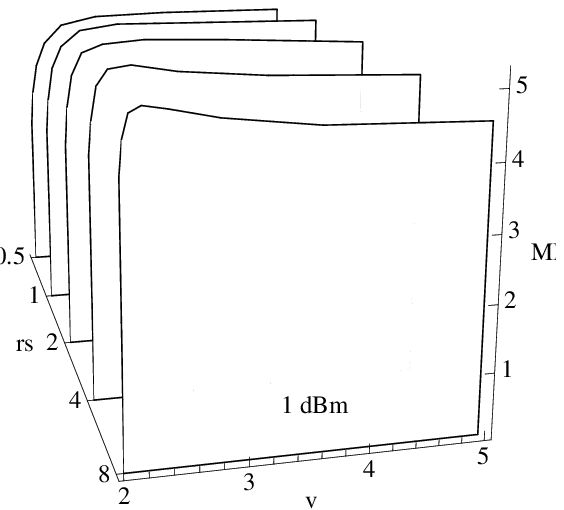} & \includegraphics[width=4cm]{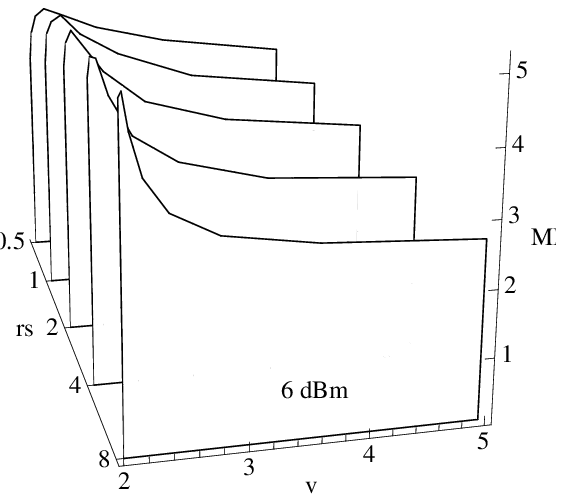}
\end{tabular}
\caption{Lower bounds on capacity from \theoref{th:clb} as a function of $\nu$, for various parameters $P$ and $r_1^2/s$. The memory is $N=1$.}
\label{fig:clb-vs-parameters}
\end{center}
\end{figure}

An attractive distribution in this context is to let the variable-amplitude symbols follow a circularly symmetric \emph{bivariate t-distribution}
 \cite[p.~86]{fang90}, \cite[p.~1]{kotz04},
\eqlab{t}{
f_{X}(x) = \frac{1}{2\pi s} \left( 1+\frac{|x|^2}{\nu s} \right)^{-(1+\nu/2)}
,}
where $X$ (with magnitude $R=\abs{X}$) denotes one such variable-amplitude symbol, $\nu$ is a shape parameter, and $s$ scales the variance, which equals \cite[p.~11]{kotz04} $\E[|X|^2] =\E[R^2]= 2\nu s/(\nu-2)$ if $\nu>2$ and is otherwise undefined.
The shape of this distribution is similar to a Gaussian, but the heaviness of the tail can be controlled via the shape parameter $\nu$: the closer $\nu$ is to $2$, the heavier tail. This is, as we shall see later, what makes it an interesting choice for nonlinear optical channels.

Again, we consider the same scenario as in \secref{Sec:pulses}, with the system parameters given in \tabref{tab:constants}. The distribution of $R=|X|$ is given by $f_R(r) = 2\pi r f_{X}(r)$, with $f_{X}$ given by \eqref{t}. The power constraint~\eqref{eq:power_normalization}, which reduces to
\eq{
P 
= \frac{1}{2N+1}\left(2N r_1^2 + \frac{2\nu s}{\nu-2} \right)
,}
leaves two degrees of freedom to optimize for each $P$, which we can take to be the shape parameter $\nu$ and the ratio $r_1^2/s$.

\begin{figure}
\begin{center}
\newcommand{\scale}{0.85}
\psfrag{N=0}[cl][cl][\scale]{AWGN}%
\psfrag{ZERO}[cl][cl][\scale]{$N=0$}%
\psfrag{ONE}[cl][cl][\scale]{$N=1$}%
\psfrag{LowerBound}[cl][cl][\scale]{Lower bounds}%
\psfrag{OTH}[cc][cc][\scale]{$N=2,5,10,20,50$}%
\psfrag{N=infinityyyyyyyyy}[cl][cl][\scale]{GN model}%
\psfrag{xlabel}[cc][cB][\scale]{$\Px$~[dBm]}%
\psfrag{ylabel}[cc][cB][\scale]{$C$~[bit/symbol]}
\includegraphics{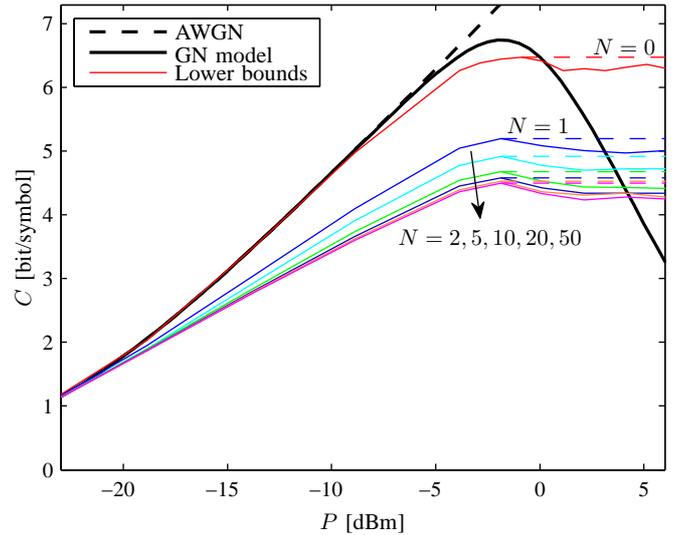}	%
\caption{Lower bounds from Theorem~\ref{th:clb} on the capacity of the finite-memory model for different values of $N$. The exact capacities of the AWGN channel and the GN model in \eqref{eq:cawgn} are included for comparison. Observe that the capacity of the finite-memory model does not converge to the capacity of the GN model as the memory $N$ increases. Dashed lines indicate improved lower bounds via the law of monotonic channel capacity.}
\label{fig:clb}
\end{center}
\end{figure}

The lower bound on the capacity of the finite-memory model given by Theorem~\ref{th:clb} is shown in \figref{fig:clb-vs-parameters} as a function of $P$, $\nu$, and $r_1^2/s$, for the special case $N=1$. The expectation in \eqref{eq:clb} was estimated by Monte Carlo integration. It can be seen that as the transmit power $P$ increases, the optimum shape parameter $\nu$ gets closer and closer to $2$. In other words, the tail gets heavier, so that at high power, it consumes almost all power, while the probability of transmitting a high amplitude $R$ is still small. In this sense, a t-distribution with a shape parameter near $2$ is similar to a satellite constellation \cite{agrell12ipc}.

Selecting the optimum parameters $\nu$ and $r_1^2/s$ for every power $P$, the capacity bound is plotted in \figref{fig:clb} as a function of transmit power $P$, for selected values of the channel memory $N$. The figure also shows the AWGN channel capacity and the exact capacity of the GN model given by \eqref{eq:cawgn}. In the linear regime, the capacity bound is close to the AWGN capacity if $N=0$, because the t-distribution is, at high values of $\nu$, approximately equal to the capacity-achieving Gaussian distribution. As $N$ increases, the capacity bound tends, still in the linear regime, to the mutual information of constant-amplitude transmission \cite{blachman53, ho02}.

Interestingly, we can see that as $N$ increases, the curves approach an asymptotic bound (the curves for $N=10$, $20$, and $50$ almost overlap). It follows that reliable communication in the high input power regime is indeed possible for every finite $N$. This result should be compared with the regular GN model, whose capacity \eqref{eq:cawgn} decreases to zero at high average transmit power \cite{bosco11}. It may seem contradictory that the GN model, which can be characterized as a limiting case of the finite-memory model (cf.~\eqref{finite.memory.IO} and \eqref{infinite.memory.IO}--\eqref{P.statistical}), nevertheless exhibits a fundamentally different channel capacity. 
This can be intuitively understood as follows. For every block of $2N+1$ symbols, we transmit $2N$ constant-amplitude symbols with low power and only one symbol with variable (potentially very large) power. 
Although the amplitude of this variable-power symbol is chosen so that the average power constraint is satisfied according to~\eqref{eq:power_normalization} (which requires averaging across many blocks of length $2N+1$), the convergence to average power illustrated in~\eqref{P.statistical} does not occur \emph{within} a block, even when $N$ is taken very large.

It can be observed that the lower bounds in \figref{fig:clb} all exhibit a low peak, before they converge to their asymptotic values at high $P$. Such bounds can always be improved using the \emph{law of monotonic channel capacity} \cite{agrell12arxiv}. Cast in the framework of this paper, this law states that the channel capacity never decreases with power for any finite-memory channel. This law does not give a capacity lower bound \emph{per se,} but it provides an instrument by which a lower bound at a certain power $P$ can be propagated to any power greater than $P$. Hence, the part of the curves in \figref{fig:clb} to the right of the peaks can be lifted up to the level of the peaks, which would yield a marginally tighter lower bounds (dashed lines in \figref{fig:clb}).

\section{Discussion and Conclusions} \label{Sec:Conclusions}

We extended the popular GN model for nonlinear fiber channels with a parameter to account for the channel memory. The extended channel model, which is given by \eqref{finite.memory.IO}, is able to model the time-varying output of an optical fiber whose input is a nonstationary process. If the input varies on a time scale comparable to or longer than the memory of the channel, then this model gives more realistic results than the regular GN model, as we showed in \figref{pulsed_tx_example}.

The validity of the GN model remains undisputed in the case of i.i.d.~input symbols, such as in an uncoded scenario with a fixed, not too heavy-tailed modulation format\footnote{Examples of ``heavy-tailed'' modulation formats are t-distributions (\secref{Sec:Capacity:Res}) and satellite constellations \cite{agrell12ipc}.} and a fixed transmit power. These are the conditions under which the GN model was derived and validated. The uncoded bit and symbol error rates computed in Sec.~\ref{Sec:Error} confirm that the finite-memory model behaves similarly to the GN model as the channel memory $N$ increases.

The scene changes completely if we instead study capacity, as in Fig.~\ref{fig:clb}. In this case, the finite-memory GN model does not, even at high $N$, behave as the regular GN model. This is because the channel capacity by definition involves a maximization over all possible transmission schemes, including nonstationary input, heavy-tailed modulation formats, etc. In the nonlinear regime, it turns out to be beneficial to transmit using a heavy-tailed
input sequence, whose output the GN model cannot reliably predict. Hence, the GN model and other infinite-memory models (in the sense defined in \secref{Sec:Model.Infinite}) should be used with caution in capacity analysis. It is still possible (and often easy) to calculate the capacity of such channel models, but this capacity should not be interpreted as the capacity of some underlying physical phenomenon with a finite memory. As a rule of thumb, if the model depends on the average transmit power, we recommend to avoid it in capacity analysis.

A challenging area for future work would be to derive more realistic finite-memory models than \eqref{finite.memory.IO}, i.e., discrete-time channel models that give the channel output as a function of a finite number of input symbols, ideally including not only a time-varying sequence of symbols but also symbols in other wavelengths, polarizations, modes, and/or cores, and to analyze these models from an information-theoretic perspective. This may lead to innovative new transmission techniques, which may potentially increase the capacity significantly over known results in the nonlinear regime. The so-called nonlinear Shannon limit, which has only been derived for infinite-memory channel models, does not prevent the existence of such techniques.

\appendices

\section{Proof of \theoref{BEP.16QAM.theorem}}\label{BEP.16QAM.theorem.Proof}

Let $\{B_q\}$, $q=1,\dots,4$, be the four bits associated with the $16$-QAM constellation point chosen as the $k$th transmitted symbol $X_k$. The BER for the 16-QAM constellation in \figref{16QAM} is given by
\begin{align}
\label{bep.proof.1}
\BEP 		& \triangleq \frac{1}{4}\sum_{q=1}^{4}\Pr\set{\hat{B}_{q} \neq {B}_{q}}\\
\label{bep.proof.2}
			& = \frac{1}{64}\sum_{q=1}^{4}\sum_{i=1}^{16}P_{\hat{B}_{q}|X_k}\bigl(\ov{B}_{q}|s_{i}\bigr),
\end{align}
where $\hat{B}_{q}$ is the estimated bit obtained by the MED detector in \eqref{med.Voronoi} and $\ov{B}$ denotes bit negation. Using the law of total probability we can then express \eqref{bep.proof.2} as
\begin{align}
\nonumber
\BEP 		& = \frac{1}{64}\sum_{q=1}^{4}\sum_{i=1}^{16}\sum_{\|\xmem\|^{2}}P_{\|\Xmem\|^{2}}\bigl(\|\xmem\|^{2}\bigr)\\
\label{bep.proof.3}
			&  \qquad\qquad \cd P_{\hat{B}_{q}|X_k,\|\Xmem\|^{2}}\bigl(\ov{B}_{q}|s_{i},\|\xmem\|^{2}\bigr).
\end{align}

We now compute the PMF  $P_{\|\Xmem\|^{2}}$. As $\|\Xmem\|^{2}$ is a sum of $2N$ i.i.d.~random variables, its PMF is the $2N$-fold self-convolution of the PMF of one such random variable. This convolution can be readily computed using probability generating functions~\cite[Sec.~5.1]{grimmett01-a}.
Let 
\begin{align}
\nonumber
\hat{P}_{|X_k|^2}(z)	&=  \frac{1}{4}(z^{2\Delta^{2}}+2z^{10\Delta^{2}}+z^{18\Delta^{2}})\\
				& =  \frac{1}{4}\bigl(z^{\Delta^{2}}+z^{9\Delta^{2}}\bigr)^{2}\label{eq:pgf.single.variable}
\end{align}
denote the probability generating function of $|X_k|^2$.
The probability generating function of $\|\Xmem\|^{2}$ is given by
\begin{align}
\label{bep.proof.gen.1}
\hat{P}_{\|\Xmem\|^{2}}(z) & = \bigl(\hat{P}_{|X_k|^{2}}(z)\bigr)^{2N}\\
\label{bep.proof.gen.2}
						& = \frac{1}{4^{2N}}\bigl(z^{\Delta^{2}}+z^{9\Delta^{2}}\bigr)^{4N}\\
\label{bep.proof.gen.3}
						& = \sum_{l=0}^{4N}\frac{1}{4^{2N}}\nchoosek{4N}{l}z^{(4N+8l)\Delta^{2}}.
\end{align}
We see from~\eqref{bep.proof.gen.3} that the possible outcomes of $\|\Xmem\|^{2}$ are 
\begin{align}\label{delta.l}
\delta_{l}\triangleq {(4N+8l)\Delta^{2}}, \quad l=0,1,\ld,4N, 
\end{align}
and $\|\Xmem\|^{2} = \delta_l$ occurs with probability ${\nchoosek{4N}{l}}{4^{-2N}}$. Using this in \eqref{bep.proof.3} yields
\begin{align}
\notag
\BEP 		& = \frac{4^{-3}}{4^{2N}}\suml\nchoosek{4N}{l}\sum_{q=1}^{4}\sum_{i=1}^{16}P_{\hat{B}_{q} |X_k,\|\Xmem\|^{2}} (\ov{B}_{q}|s_{i},\delta_{l})
\\
\nonumber
			& = \frac{4^{-3}}{4^{2N}}\suml\nchoosek{4N}{l}\sum_{q=1}^{4}\sum_{i=1}^{16}\sum_{\substack{j=1\\c_{j,q}\neq c_{i,q}}}^{16}\\
\label{bep.proof.5}
			& \qquad\, \int_{\mcV_{j}}\frac{\pi^{-1}}{\Szsilowerl}\exp{\biggl(-\frac{|\y-s_{i}|^{2}}{\Szsilowerl}\biggr)}\,\tr{d}y,
\end{align}
where \eqref{bep.proof.5} follows from \eqref{ctp} and $c_{j,q}$ represents the $q$th bit label of the symbol $s_{j}$ for $j=1,\ld,16$. 

The density in the integral in \eqref{bep.proof.5} corresponds to a Gaussian random variable with total variance $\Szsilowerl$, and thus, we now focus on the function $\Szsilowerl$. First, we express the constellation points indices as $\set{1,2,\ld,16}=\mcI_{1}\cup \mcI_{5}\cup \mcI_{9}$, where $\mcI_{1}\triangleq\set{6,7,10,11}$, $\mcI_{5}\triangleq\set{2,3,5,8,9,12,14,15}$, and $\mcI_{9}\triangleq\set{1,4,13,16}$. From \figref{16QAM}, we see that $|s_{i}|^{2}=2\Delta^{2}$ if $i\in\mcI_{1}$, $|s_{i}|^{2}=10\Delta^{2}$ if $i\in\mcI_{5}$, and $|s_{i}|^{2}=18\Delta^{2}$ if $i\in\mcI_{9}$. Using the definition of $\Szsilowerl$ in \eqref{noise.variance} together with \eqref{delta.l} and $\Px=10\Delta^{2}$, we obtain
\begin{align}\label{noise.variance.cases}
&\Szsilowerl
=\notag\\
&\quad
\begin{cases}
\PASE+\frac{\eta}{(2N+1)^{3}}\left(\frac{\Px(2N+4l+1)}{5}\right)^{3}\!\!,& \tnr{if $i\in\mcI_{1}$}\\
\PASE+\frac{\eta}{(2N+1)^{3}}\left(\frac{\Px(2N+4l+5)}{5}\right)^{3}\!\!,& \tnr{if $i\in\mcI_{5}$}\\
\PASE+\frac{\eta}{(2N+1)^{3}}\left(\frac{\Px(2N+4l+9)}{5}\right)^{3}\!\!,& \tnr{if $i\in\mcI_{9}$}\\
\end{cases}.
\end{align}
We recognize the three values of $\Szsilowerl$ in \eqref{noise.variance.cases} as $\gammaltN{l}{1}{N}$, $\gammaltN{l}{5}{N}$, and $\gammaltN{l}{9}{N}$, respectively. Combining this with $\Px=10\Delta^{2}$ and inspecting the constellation and labeling in \figref{16QAM} yields \eqref{BEP.16QAM}.

\section{Proof of \theoref{SEP.16QAM.theorem}}\label{SEP.16QAM.theorem.Proof}

The SER for the 16-QAM constellation in \figref{16QAM} is
\begin{align}
\label{sep}
\SEP 		& \triangleq \Pr\set{\hX_k \neq \X_k}\\
\label{sep.proof.1}
			& = \frac{1}{16}\sumi\Pr\set{\hX_k \neq s_i|\X_k=s_i}\\
\label{sep.proof.2}
			& = \frac{1}{16}\sumi\sum_{\substack{j=1\\j\neq i}}^{16}\Pr\set{Y_k \in \mcV_{j}|\X_k=s_i}.
\end{align}
By conditioning on the possible values of $\|\Xmem\|^{2}$, we obtain
\begin{align}
\nonumber
\SEP 		& = \frac{4^{-2}}{4^{2N}}\suml\nchoosek{4N}{l}\sumi\sum_{\substack{j=1\\j\neq i}}^{16}\\
\label{sep.proof.3}
			&\quad\qquad\quad\Pr\set{Y_k \in \mcV_{j} |\X_k=s_i,\|\Xmem\|^{2}=\delta_{l}}\\
\nonumber
			& =\frac{4^{-2}}{4^{2N}}\suml\nchoosek{4N}{l}\sumi\sum_{\substack{j=1\\j\neq i}}^{16}\\
\label{sep.proof.4}
			& \qquad\qquad \int_{\mcV_{j}}\frac{\pi^{-1}}{\Szsilowerl}\exp{\biggl(-\frac{|\y-s_{i}|^{2}}{\Szsilowerl}\biggr)}\,\tr{d}y,
\end{align}
where $\delta_{l}$ is given by \eqref{delta.l}.

The expression in \eqref{SEP.16QAM} is obtained by recognizing the density in the integral in \eqref{sep.proof.4} as a Gaussian random variable with total variance given by \eqref{noise.variance.cases}, and by integrating, for each $i$ in the sets $\mcI_{1}$, $\mcI_{5}$, and $\mcI_{9}$, over $\mcV_{j}$ with $j\neq i$. This completes the proof of \theoref{SEP.16QAM.theorem}.

\section{Proof of Corollary \ref{16QAM.infmemory.theorem}}\label{16QAM.infmemory.theorem.Proof}
 
The SER in \eqref{SEP.16QAM} can be expressed as 
\begin{align}
\label{SEP.16QAM.u}
\SEP & = \frac{1}{4}
\sum_{\substack{e\in\set{1,2}\\t\in\set{1,5,9}}}S_{e,t}
\suml \frac{1}{4^{2N}}\nchoosek{4N}{l}
u_{e}\biggl(\frac{4l+t-1}{4(2N+1)}\biggr),
\end{align}
where 
\eqlab{u.def}{
u_{e}&(x) \triangleq
\QF\lefto(\sqrt{\frac{\Px/5}{\PASE+({1+4x})^{3}\eta \left({\Px}/{5}\right)^{3}}}\right)^{e}}
is a continuous and bounded function in $[0,2]$ for any $e\in\set{1,2}$ and $t\in\set{1,5,9}$. We can interpret the innermost sum in \eqref{SEP.16QAM.u} in probabilistic terms as 
\begin{align}
\nonumber
\suml \frac{1}{4^{2N}} \nchoosek{4N}{l} u_{e} & \biggl(\frac{4l+t-1}{4(2N+1)}\biggr) \\
\label{eq:expectation.binomial}
&=\Ex\lefto[u_{e}\biggl(\frac{4S_{4N}+t-1}{4(2N+1)}\biggr)\right],
\end{align}
where $S_{4N}$ is a binomial random variable with parameters $(4N,1/2)$, i.e., $S_{4N}$ is the sum of $4N$ i.i.d.\ Bernoulli random variables that take values $0$ and $1$ with the same probability. We use the notation $S_{4N}$ to emphasize the dependency on $N$. To establish~\eqref{SEP.16QAM.infmemory}, we first calculate
\begin{align}
\nonumber
   \lim_{N\to \infty}\Ex&\lefto[u_{e}\biggl(\frac{4S_{4N}+t-1}{4(2N+1)}\biggr)\right] \\
\label{eq:limit.binomial.0}
   &=\Ex\lefto[ \lim_{N\to \infty} u_{e}\biggl(\frac{4S_{4N}+t-1}{4(2N+1)}\biggr)\right] \\
\label{eq:limit.binomial.1}
   &= \Ex\lefto[  u_{e}\biggl(\lim_{N\to \infty} \frac{4S_{4N}+t-1}{4(2N+1)}\biggr)\right] \\
   &=\Ex[u_{e}(1)] \label{eq:limit.binomial} \\
   &= u_{e}(1).
\end{align}
Here, \eqref{eq:limit.binomial.0} follows from the dominated convergence theorem~\cite[Sec.~5.6.(12).(b)]{grimmett01-a}, whose application is possible because $u_{e}(x)$ is a bounded function, \eqref{eq:limit.binomial.1} holds because $u_{e}(x)$ is continuous, and \eqref{eq:limit.binomial} follows from the law of large numbers (see e.g.,~\cite[Sec.~7.4.(3)]{grimmett01-a}). The proof of \eqref{SEP.16QAM.infmemory} is completed by using
\eqlab{u.1}{
u_{e}&(1) = \QF\lefto(\sqrt{\frac{\Px/5}{\PASE+\eta\Px^{3}}}\right)^{e}
}
and \eqref{eq:limit.binomial} in \eqref{SEP.16QAM.u} together with \eqref{S.1}--\eqref{S.9}.

The proof of the BER expression in~\eqref{BEP.16QAM.infmemory} follows steps similar to the ones we presented above.

\section{Proof of Theorem \ref{th:clb}}\label{app:clb}
Consider a sequence of independent symbols $X_k = R_k e^{\jmath \Phi_k}, k\in \mathbb{Z}$, where for each $k$, the magnitude $R_k$ is independent of the phase $\Phi_k$, which is uniform in $[0,2\pi)$. 
The magnitude $R_k$ is distributed according to $f_R$ if $k=0 \mod (2N+1)$ and is otherwise equal to the constant $r_1$. 
Furthermore, $f_R$ and $r_1$ are chosen so that~\eqref{eq:power_normalization} holds, which guarantees that the average power constraint is satisfied.
We will next show that the right-hand side of \eqref{eq:clb} is the mutual information (in bits per channel use) obtainable with this input distribution.
Hence, it is a lower bound on capacity.

We define blocks of length $2N+1$ of transmitted and received symbols as
\eq{
\bY_l &\triangleq \bY_{l(2N+1)-N}^{l(2N+1)+N}, \\
\bX_l &\triangleq \bX_{l(2N+1)-N}^{l(2N+1)+N}
}
for $l\in\mathbb{Z}$. 
Let us focus for a moment on the received block $\bY_0$.
Let $Y_k$ be the $k$th element ($k=-N,\dots,N$) of $\bY_0$.
It follows from~\eqref{finite.memory.IO} that the additive noise contribution to $Y_k$ depends on the input vector  $\|\bX_{k-N}^{k+N}\|$, which may span more than one input block.
By construction, however, all elements of $\bX_{k-N}^{k+N}$ with the exception of $X_0$ have constant magnitude equal to $r_1$.
Hence,
\begin{IEEEeqnarray}{rCL}
  \|\bX_{k-N}^{k+N}\|^2=\abs{X_0}^2+2Nr_1^{2}.
\end{IEEEeqnarray}
This implies that
\begin{multline}\label{eq:pdf.two.level.dist}
  f_{Y_k | \bX_{k-N}^{k+N}}(y_k | \bx_{k-N}^{k+N})\\=\frac{1}{\pi \rho(2Nr^2_1+\abs{x_0}^2)}\exp\left(\frac{\abs{y_k-x_k}^2}{\rho(2Nr^2_1+\abs{x_0}^2)}\right).
\end{multline}
We see from~\eqref{eq:pdf.two.level.dist} that each output sample $Y_k$ in $\bY_0$ actually depends on the input symbols only through $X_k$ and $X_0$. 
We then conclude that $\bY_0$ depends on the whole input sequence only through $\bX_0$.
But this, together with the assumption of independent input  symbols, implies that the output blocks $\{\bY_l\}$ are independent. 
Hence, from \eqref{eq:lbmemory},
\eqlab{lbblock}{
C \ge \frac{1}{2N+1} I(\bX_l,;\bY_l)
}
for an arbitrary $l\in\mathbb{Z}$, say, $l=0$. 

Next, we calculate $I(\bX_0;\bY_0)$.
The mutual information can be decomposed into differential entropies as
\eqlab{ixy}{
I(\bX_0;\bY_0) = h(\bY_0) - h(\bY_0|\bX_0)
,}
where 
\eqlab{hy}{
h(\bY_0) &= -\E[\log_2 f_{\bY_0}(\bY_0)], \\
h(\bY_0|\bX_0) &= -\E[\log_2 f_{\bY_0|\bX_0}(\bY_0|\bX_0)] \label{hyx}
.}

We start by evaluating~\eqref{hyx}. 
Because of~\eqref{eq:pdf.two.level.dist}, the conditional distribution of $\bY_0$ given $\bX_0$ is the multivariate Gaussian density
\begin{multline}
f_{\bY_0|\bX_0}(\by_0|\bx_0)\\=\frac{1}{ \bigl(\pi\rho(2Nr_1^2+\abs{x_0}^2)\bigr)^{2N+1}}\exp\lefto(-\frac{\|\by_0-\bx_0\|^2}{\rho(2Nr^2_1+\abs{x_0}^2)}\right). 
\end{multline}
Using~\cite[Theorem~8.4.1]{cover06}, we conclude that
\begin{IEEEeqnarray}{rCL}\label{eq:cond_diff_ent}
  h(\bY_0|\bX_0)=(2N+1)\E[\log_2\pi \rho(2Nr_1^2+\abs{X_0}^2)],
\end{IEEEeqnarray}
where the expectation is with respect to the random variable $\abs{X_0}$, which is distributed according to $f_R$.

To evaluate \eqref{hy}, we start by noting that all elements of $\bY_0$ have uniform phase because the transmitted symbols and the additive noise samples have uniform phase by assumption. 
We use this property to simplify~\eqref{hy}.
Specifically, let $U_k=\abs{Y_k}^2$ and
\begin{IEEEeqnarray}{rCL}
  \bU\triangleq [U_{-N},U_{-N+1},\ldots,U_{N}].
\end{IEEEeqnarray}
By~\cite[eq.~(320)]{lapidoth03-10a}
\begin{IEEEeqnarray}{rCL}
  h(\bY_0)=(2N+1)\log_2 (\pi) + h(\bU).
\end{IEEEeqnarray}
To evaluate $h(\bU)=-\E[\log_2(f_{\bU}(\bU))]$, we first derive the conditional distribution $f_{\bU | \abs{X_0}}$ of $\bU$ given $\abs{X_0}$.
Note that $U_k$ has the same distribution as $\abs{\abs{X_k}+\sqrt{\rho(2Nr^2_1+\abs{X_0}^2)}\tilde{Z}_k }^2$ (see~\eqref{eq:awgn} and~\eqref{finite.memory.IO}).
Hence, given $\abs{X_0}=r$, the random variables $\{2U_k/\rho(2Nr^2_1+r^2)\}$ follow a noncentral chi-square distribution with two degrees of freedom and noncentrality parameters $\{2\abs{X_k}^2/\rho(2Nr^2_1+r^2)\}$, where $\abs{X_k}=r_1$ if $k\neq 0$ and $\abs{X_k}=r$ otherwise.
Furthermore, these random variables are conditionally independent given $\abs{X_0}$.
Using the change of variable theorem for transformation of random variables, we finally obtain after algebraic manipulations 
\begin{IEEEeqnarray}{rCL}\label{eq:pdf_cond}
  f_{\bU|\abs{X_0}}(\bu | r) &=&\frac{\exp\lefto(-\frac{\sum_{k=-N}^N u_k+2Nr_1^2+r^2 }{\rho(2Nr^2_1+r^2)}\right)}{\bigl(\rho(2Nr^2_1+r^2)\bigr)^{2N+1}}\notag\\
  &&\cdot I_0\lefto( \frac{2r\sqrt{u_0}}{\rho(2N r_1^2+r^2)}\right) \notag\\
  &&\cdot \prod_{\substack{k=-N\\k\ne0}}^{N} I_0\lefto( \frac{2r_1 \sqrt{u_k}}{\rho(2Nr_1^2+r^2)}\right).
\end{IEEEeqnarray}
The probability distribution $f_{\bU}$, which is given in~\eqref{fy.th.clb}, is obtained from~\eqref{eq:pdf_cond} by taking the expectation with respect to$f_R$, the probability distribution of $\abs{X_0}$.
Finally, we obtain the capacity lower bound~\eqref{eq:clb} by substituting~\eqref{fy.th.clb} into~\eqref{hy} and~\eqref{eq:cond_diff_ent} into~\eqref{hyx}, by computing the difference between the two resulting differential entropies according to~\eqref{ixy}, and by dividing by $2N+1$.

\balance


\begin{thebibliography}{10}
\providecommand{\url}[1]{#1}
\csname url@samestyle\endcsname
\providecommand{\newblock}{\relax}
\providecommand{\bibinfo}[2]{#2}
\providecommand{\BIBentrySTDinterwordspacing}{\spaceskip=0pt\relax}
\providecommand{\BIBentryALTinterwordstretchfactor}{4}
\providecommand{\BIBentryALTinterwordspacing}{\spaceskip=\fontdimen2\font plus
\BIBentryALTinterwordstretchfactor\fontdimen3\font minus
  \fontdimen4\font\relax}
\providecommand{\BIBforeignlanguage}[2]{{%
\expandafter\ifx\csname l@#1\endcsname\relax
\typeout{** WARNING: IEEEtran.bst: No hyphenation pattern has been}%
\typeout{** loaded for the language `#1'. Using the pattern for}%
\typeout{** the default language instead.}%
\else
\language=\csname l@#1\endcsname
\fi
#2}}
\providecommand{\BIBdecl}{\relax}
\BIBdecl

\bibitem{sun08}
H.~Sun, K.-T. Wu, and K.~Roberts, ``Real-time measurements of a 40 {Gb/s}
  coherent system,'' \emph{Opt. Express}, vol.~16, no.~2, pp. 873--879, 2008.

\bibitem{roberts09}
K.~Roberts, M.~O'Sullivan, K.-T. Wu, H.~Sun, A.~Awadalla, D.~J. Krause, and
  C.~Laperle, ``{Performance of dual-polarization QPSK for optical transport
  systems},'' \emph{J. Lightw. Technol.}, vol.~27, no.~16, pp. 3546--3559, Aug.
  2009.

\bibitem{Ellis10}
A.~D. Ellis, J.~Zhao, and D.~Cotter, ``Approaching the non-linear {S}hannon
  limit,'' \emph{J. Lightw. Technol.}, vol.~28, no.~4, pp. 423--433, Feb. 2010.

\bibitem{Mecozzi12}
A.~Mecozzi and R.-J. Essiambre, ``Nonlinear {S}hannon limit in pseudolinear
  coherent systems,'' \emph{J. Lightw. Technol.}, vol.~30, no.~12, pp.
  2011--2024, June 2012.

\bibitem{splett93}
A.~Splett, C.~Kurtzke, and K.~Petermann, ``Ultimate transmission capacity of
  amplified optical fiber communication systems taking into account fiber
  nonlinearities,'' in \emph{Proc. European Conference on Optical Communication
  (ECOC)}, Montreux, Switzerland, Sept. 1993.

\bibitem{tang02}
J.~Tang, ``{The channel capacity of a multispan DWDM system employing
  dispersive nonlinear optical fibers and an ideal coherent optical
  receiver},'' \emph{J. Lightw. Technol.}, vol.~20, no.~7, pp. 1095--1101, July
  2002.

\bibitem{poggiolini11}
P.~Poggiolini, A.~Carena, V.~Curri, G.~Bosco, and F.~Forghieri, ``Analytical
  modeling of nonlinear propagation in uncompensated optical transmission
  links,'' \emph{{IEEE} Photon. Technol. Lett.}, vol.~23, no.~11, pp. 742--744,
  June 2011.

\bibitem{bosco11}
G.~Bosco, P.~Poggiolini, A.~Carena, V.~Curri, and F.~Forghieri, ``{Analytical
  results on channel capacity in uncompensated optical links with coherent
  detection},'' \emph{Opt. Express}, vol.~19, no.~26, pp. B440--B449, Dec.
  2011.

\bibitem{carena12}
A.~Carena, V.~Curri, G.~Bosco, P.~Poggiolini, and F.~Forghieri, ``Modeling of
  the impact of nonlinear propagation effects in uncompensated optical coherent
  transmission links,'' \emph{J. Lightw. Technol.}, vol.~30, no.~10, pp.
  1524--1539, May 2012.

\bibitem{poggiolini12}
P.~Poggiolini, ``The {GN} model of non-linear propagation in uncompensated
  coherent optical systems,'' \emph{J. Lightw. Technol.}, vol.~24, no.~30, pp.
  3875--3879, Dec. 2012.

\bibitem{beygi12}
L.~Beygi, E.~Agrell, P.~Johannisson, M.~Karlsson, and H.~Wymeersch, ``A
  discrete-time model for uncompensated single-channel fiber-optical links,''
  \emph{{IEEE} Trans. Commun.}, vol.~60, no.~11, pp. 3440--3450, Nov. 2012.

\bibitem{Johannisson13}
P.~Johannisson and M.~Karlsson, ``Perturbation analysis of nonlinear
  propagation in a strongly dispersive optical communication system,'' \emph{J.
  Lightw. Technol.}, vol.~31, no.~8, pp. 1273--1282, Apr. 2013.

\bibitem{grellier11}
E.~Grellier and A.~Bononi, ``{Quality parameter for coherent transmissions with
  Gaussian-distributed nonlinear noise},'' \emph{Opt. Express}, vol.~19,
  no.~13, pp. 12\,781--12\,788, June 2011.

\bibitem{beygi13}
L.~Beygi, N.~V. Irukulapati, E.~Agrell, P.~Johannisson, M.~Karlsson,
  H.~Wymeersch, P.~Serena, and A.~Bononi, ``On nonlinearly-induced noise in
  single-channel optical links with digital backpropagation,'' \emph{Optics
  Express}, vol.~21, no.~22, pp. 26\,376--26\,386, Nov. 2013.

\bibitem{vacondio12}
F.~V.~O. Rival, C.~Simonneau, E.~Grellier, A.~Bononi, L.~Lorcy, J.-C. Antona,
  and S.~Bigo, ``On nonlinear distortions of highly dispersive optical coherent
  systems,'' \emph{Opt. Express}, vol.~20, no.~2, pp. 1022--1032, Jan. 2012.

\bibitem{shannon48}
C.~E. Shannon, ``A mathematical theory of communication,'' \emph{Bell System
  Technical Journal}, vol.~27, pp. 379--423, 623--656, July, Oct. 1948.

\bibitem{cover06}
T.~M. Cover and J.~A. Thomas, \emph{Elements of Information Theory},
  2nd~ed.\hskip 1em plus 0.5em minus 0.4em\relax Hoboken, NJ: Wiley, 2006.

\bibitem{Stark99}
J.~B. Stark, ``Fundamental limits of information capacity for optical
  communications channels,'' in \emph{Proc. European Conference on Optical
  Communication (ECOC)}, Nice, France, Sep. 1999.

\bibitem{mitra01}
P.~P. Mitra and J.~B. Stark, ``Nonlinear limits to the information capacity of
  optical fibre communications,'' \emph{Nature}, vol. 411, pp. 1027--1030, June
  2001.

\bibitem{turitsyn03}
K.~S. Turitsyn, S.~A. Derevyanko, I.~V. Yurkevich, and S.~K. Turitsyn,
  ``Information capacity of optical fiber channels with zero average
  dispersion,'' \emph{Physical Review Letters}, vol.~91, no.~20, pp.
  203\,901\raisebox{.44ex}{\rule{.27em}{.14ex}}1--4, Nov. 2003.

\bibitem{djordjevic05}
I.~B. Djordjevic and B.~Vasic, ``Achievable information rates for high-speed
  long-haul optical transmission,'' \emph{J. Lightw. Technol.}, vol.~11,
  no.~23, pp. 3755--3763, Nov. 2005.

\bibitem{taghavi06}
M.~H. Taghavi, G.~C. Papen, and P.~H. Siegel, ``On the multiuser capacity of
  {WDM} in a nonlinear optical fiber: Coherent communication,'' \emph{{IEEE}
  Trans. Inf. Theory}, vol.~52, no.~11, pp. 5008--5022, Nov. 2006.

\bibitem{Essiambre10}
R.-J. Essiambre, G.~Kramer, P.~J. Winzer, G.~J. Foschini, and B.~Goebel,
  ``Capacity limits of optical fiber networks,'' \emph{J. Lightw. Technol.},
  vol.~28, no.~4, pp. 662--701, Feb. 2010.

\bibitem{secondini13}
M.~Secondini, E.~Forestieri, and G.~Prati, ``Achievable information rate in
  nonlinear {WDM} fiber-optic systems with arbitrary modulation formats and
  dispersion maps,'' \emph{J. Lightw. Technol.}, vol.~31, no.~23, pp.
  3839--3852, Dec. 2013.

\bibitem{Dar14}
R.~Dar, M.~Shtaif, and M.~Feder, ``New bounds on the capacity of the nonlinear
  fiber-optic channel,'' \emph{Optics Letters}, vol.~39, no.~2, pp. 398--401,
  Jan. 2014.

\bibitem{Kahn04}
J.~M. Kahn and K.-P. Ho, ``Spectral efficiency limits and modulation/detection
  techniques for {DWDM} systems,'' \emph{{IEEE} Journal of Selected Topics in
  Quantum Electronics}, vol.~10, no.~2, pp. 259--272, Mar./Apr 2004.

\bibitem{Narimanov02}
E.~Narimanov and P.~Mitra, ``The channel capacity of a fiber optics
  communication system: perturbation theory,'' \emph{J. Lightw. Technol.},
  vol.~20, no.~3, pp. 530--537, Mar. 2002.

\bibitem{wegener04}
L.~G.~L. Wegener, M.~L. Povinelli, A.~G. Green, P.~P. Mitra, J.~B. Stark, and
  P.~B. Littlewood, ``The effect of propagation nonlinearities on the
  information capacity of {WDM} optical fiber systems: Cross-phase modulation
  and four-wave mixing,'' \emph{Physica D: Nonlinear Phenomena}, vol. 189, no.
  1-2, pp. 81--99, Feb. 2004.

\bibitem{essiambre08}
R.-J. Essiambre, G.~J. Foschini, G.~Kramer, and P.~J. Winzer, ``Capacity limits
  of information transport in fiber-optic networks,'' \emph{Physical Review
  Letters}, vol. 101, no.~16, pp.
  163\,901\raisebox{.44ex}{\rule{.27em}{.14ex}}1--4, Oct. 2008.

\bibitem{freckmann09}
T.~Freckmann, R.-J. Essiambre, P.~J. Winzer, G.~J. Foschini, and G.~Kramer,
  ``Fiber capacity limits with optimized ring constellations,'' \emph{{IEEE}
  Photon. Technol. Lett.}, vol.~21, no.~20, pp. 1496--1498, Oct. 2009.

\bibitem{djordjevic10}
I.~B. Djordjevic, H.~G. Batshon, L.~Xu, and T.~Wang, ``Coded
  polarization-multiplexed iterative polar modulation ({PM-IPM}) for beyond 400
  {G}b/s serial optical transmission,'' in \emph{Proc. Optical Fiber
  Communication Conference (OFC)}, San Diego, CA, Mar. 2010.

\bibitem{killey11}
R.~I. Killey and C.~Behrens, ``Shannon's theory in nonlinear systems,''
  \emph{Journal of Modern Optics}, vol.~58, no.~1, pp. 1--10, Jan. 2011.

\bibitem{agrell09}
E.~Agrell and M.~Karlsson, ``Power-efficient modulation formats in coherent
  transmission systems,'' \emph{J. Lightw. Technol.}, vol.~27, no.~22, pp.
  5115--5126, Nov. 2009.

\bibitem{agrellofc13}
------, ``{WDM Channel Capacity and its Dependence on Multichannel Adaptation
  Models},'' in \emph{Proc. Optical Fiber Communication Conference (OFC)},
  Anaheim, CA, Mar. 2013.

\bibitem{Goebel11}
B.~Goebel, R.-J. Essiambre, G.~Kramer, P.~J. Winzer, and N.~Hanik,
  ``Calculation of mutual information for partially coherent {G}aussian
  channels with applications to fiber optics,'' \emph{{IEEE} Trans. Inf.
  Theory}, vol.~57, no.~9, pp. 5720--5736, Sep. 2011.

\bibitem{bosco12}
G.~Bosco, P.~Poggiolini, A.~Carena, V.~Curri, and F.~Forghieri, ``Analytical
  results on channel capacity in uncompensated optical links with coherent
  detection: {Erratum},'' \emph{Opt. Express}, vol.~20, no.~17, pp.
  19\,610--19\,611, Aug. 2012.

\bibitem{agrell12ipc}
E.~Agrell and M.~Karlsson, ``Satellite constellations: {Towards} the nonlinear
  channel capacity,'' in \emph{Proc. IEEE Photon. Conf. (IPC)}, Burlingame, CA,
  Sept. 2012.

\bibitem{carena10}
A.~Carena, G.~Bosco, V.~Curri, P.~Poggiolini, M.~T. Taiba, and F.~Forghieri,
  ``Statistical characterization of {PM-QPSK} signals after propagation in
  uncompensated fiber links,'' in \emph{Proc. European Conference on Optical
  Communication (ECOC)}, London, U.K., Sept. 2010.

\bibitem{koch09-08a}
T.~Koch, A.~Lapidoth, and P.~Sotiriadis, ``Channels that heat up,'' \emph{IEEE
  Trans. Inf. Theory}, vol.~55, no.~8, pp. 3594 --3612, Aug. 2009.

\bibitem{gallager68}
R.~G. Gallager, \emph{Information Theory and Reliable Communication}.\hskip 1em
  plus 0.5em minus 0.4em\relax New York, NY: Wiley, 1968.

\bibitem{ip07}
E.~Ip and J.~M. Kahn, ``Digital equalization of chromatic dispersion and
  polarization mode dispersion,'' \emph{J. Lightw. Technol.}, vol.~25, no.~8,
  pp. 2033--2043, Aug. 2007.

\bibitem{Agrell04}
E.~Agrell, J.~Lassing, E.~G. Str\"{o}m, and T.~Ottosson, ``On the optimality of
  the binary reflected {G}ray code,'' \emph{{IEEE} Trans. Inf. Theory},
  vol.~50, no.~12, pp. 3170--3182, Dec. 2004.

\bibitem{Fitz94}
M.~P. Fitz and J.~P. Seymour, ``On the bit error probability of qam
  modulation,'' \emph{International Journal of Wireless Information Networks},
  vol.~1, no.~2, pp. 131--139, Apr. 1994.

\bibitem{Simon95_Book}
M.~K. Simon, S.~M. Hinedi, and W.~C. Lindsey, \emph{Digital Communication
  Techniques: Signal Design and Detection}.\hskip 1em plus 0.5em minus
  0.4em\relax Englewood Cliffs, NJ: Prentice-Hall, 1995.

\bibitem{mecozzi94}
A.~Mecozzi, ``Limits to long-haul coherent transmission set by the kerr
  nonlinearity and noise of the in-line amplifiers,'' \emph{J. Lightw.
  Technol.}, vol.~12, no.~11, pp. 1993--2000, Nov. 1994.

\bibitem{demir07}
A.~Demir, ``Nonlinear phase noise in optical-fiber-communication systems,''
  \emph{J. Lightw. Technol.}, vol.~25, no.~8, pp. 2002--2032, Aug. 2007.

\bibitem{agrawal10}
G.~P. Agrawal, \emph{Fiber-optic communication systems}, 4th~ed.\hskip 1em plus
  0.5em minus 0.4em\relax Wiley, 2010.

\bibitem{verdu94}
S.~Verd\'u and T.~S. Han, ``A general formula for channel capacity,''
  \emph{{IEEE} Trans. Inf. Theory}, vol.~40, no.~4, pp. 1147--1157, July 1994.

\bibitem{fang90}
K.-T. Fang, S.~Kotz, and K.~W. Ng, \emph{Symmetric Multivariate and Related
  Distributions}.\hskip 1em plus 0.5em minus 0.4em\relax Springer, 1990.

\bibitem{kotz04}
S.~Kotz and S.~Nadarajah, \emph{Multivariate {t} Distributions and Their
  Applications}.\hskip 1em plus 0.5em minus 0.4em\relax Cambridge University
  Press, 2004.

\bibitem{blachman53}
N.~M. Blachman, ``A comparison of the informational capacities of amplitude-
  and phase-modulation communication systems,'' \emph{Proceedings of the
  I.R.E.}, vol.~41, no.~6, pp. 748--759, June 1953.

\bibitem{ho02}
K.-P. Ho and J.~M. Kahn, ``Channel capacity of {WDM} systems using
  constant-intensity modulation formats,'' in \emph{Proc. Optical Fiber
  Communication Conference (OFC)}, Anaheim, CA, Mar. 2002.

\bibitem{agrell12arxiv}
\BIBentryALTinterwordspacing
E.~Agrell, ``On monotonic capacity--cost functions,'' 2012, preprint. [Online].
  Available: \url{http://arxiv.org/abs/1209.2820}
\BIBentrySTDinterwordspacing

\bibitem{grimmett01-a}
G.~R. Grimmett and D.~R. Stirzaker, \emph{Probability and Random Processes},
  3rd~ed.\hskip 1em plus 0.5em minus 0.4em\relax Oxford University Press, 2001.

\bibitem{lapidoth03-10a}
A.~Lapidoth and S.~M. Moser, ``Capacity bounds via duality with applications to
  multiple-antenna systems on flat-fading channels,'' \emph{{IEEE} Trans. Inf.
  Theory}, vol.~49, no.~10, pp. 2426--2467, Oct. 2003.

\end{thebibliography}
\end{document}